\documentclass[prb,twocolumn,aps,showpacs]{revtex4-1}
\usepackage{t1enc}
\usepackage[final]{graphicx}
\usepackage{graphicx}
\usepackage{epsfig}
\usepackage{amsmath}                                                                                                      
\usepackage[normalem]{ulem}

\begin{document}

\title{Single-Chain Magnets from sharp to broad domain walls}  

\author{Orlando V. Billoni$^{1}$} 
\email{billoni@famaf.unc.edu.ar}
\author{Vivien Pianet$^{2, 3}$}
\author{Danilo Pescia$^4$} 
\author{Alessandro Vindigni$^4$}
\email{vindigni@phys.ethz.ch}
\affiliation{$^1$Facultad de Matem\'atica, Astronom\'{\i}a y F\'{\i}sica (IFEG-CONICET), Universidad Nacional de C\'ordoba, Ciudad Universitaria, 5000 C\'ordoba, Argentina}
\affiliation{$^{2}$CNRS, UPR 8641, Centre de Recherche Paul Pascal (CRPP), Equipe "Mat\'eriaux Mol\'eculaires Magn\'etiques", 115 avenue du Dr. Albert Schweitzer, Pessac, F-33600, France} 
\affiliation{$^{3}$Universit\'e de Bordeaux, UPR 8641, Pessac, F-33600, France}
\affiliation{$^4$Laboratorium f\"ur Festk\"orperphysik, Eidgen\"ossische Technische Hochschule Z\"urich, CH-8093 Z\"urich, Switzerland} 

\date{\today}                                           
                                                                                       
\begin{abstract}
We discuss time-quantified Monte-Carlo simulations on classical spin chains with uniaxial anisotropy in relation to static calculations. 
Depending on the thickness of domain walls, controlled by the relative strength of the exchange and magnetic anisotropy energy,  
we found two distinct regimes in which both the static and dynamic behavior are different. 
For broad domain walls, the interplay between localized excitations and spin waves turns out to be crucial at finite temperature. 
As a consequence, a different protocol should be followed in the experimental characterization of slow-relaxing spin chains with broad domain walls  
with respect to the usual Ising limit. 
\end{abstract}

%\pacs{75.10.Hk,75.50.Xx,75.40.Gb}
%75.10.Hk 	Classical spin models 
%75.50.Xx 	Molecular magnets 
%75.40.Gb 	Dynamic properties (dynamic susceptibility, spin waves, spin diffusion, dynamic scaling, etc.) 
\maketitle

\section{Introduction} 
The interest in the physics of domain walls (DWs) in 1d magnetic systems  
has been renewed  by the capability of controlling their motion by means of an electric current~\cite{Slonczewski_JMMM_96}.   
The technological relevance of this topic mainly derives from the possibility of employing DWs in novel magneto-storage and spintronic devices~\cite{Parkin11042008}.  
From a more fundamental point of view, the synthesis of the first slow-relaxing spin chains~\cite{Caneschi01ACIE,Caneschi02EPL,Clerac02JACS,Lescouzec03ACIE}  
gave the chance to reconsider thermally-induced DW diffusion in 1d classical spin models.    The systems displaying such a behavior  
have been named Single-Chain Magnets (SCMs), by analogy with Single-Molecule Magnets (SMMs)~\cite{Gatteschi-Sessoli_rev_SMM_03}:  
both classes of materials show a magnetic hysteresis at finite temperature due to slow dynamics rather than 
cooperative 3d ordering. Thanks to the remarkable property of being bistable at a molecular level,  
SMMs and SCMs have been proposed as possible magnetic storage units. However, the advantage of having 
identical units whose arrangement might -- in principle -- be tailored with chemical methods is counterbalanced by the relatively poor thermal stability. 
In fact, the relaxation time becomes macroscopic only at temperatures of the order of one Kelvin or lower. 
The quest to improve thermal stability of molecular magnets has called for a better understanding of their physical properties. 
The investigation and development of SCMs has become an independent and very active field of research during the last decade~\cite{Miyasaka_review,Coulon06Springer,Bogani_JMC_08}. 
At odds with SMMs, which can be considered 0d magnetic systems, SCMs develop short-range order over a distance comparable to the correlation length, $\xi$. 
The latter is expected to diverge when the temperature approaches zero. However, defects and lattice dislocations -- whose role is particularly dramatic in 1d -- typically hinder the divergence of $\xi$ below a certain temperature.   
The physics of SCMs is thus different depending on whether the correlation length exceeds or not the average size of connected spin centers. 
Finite-size effects will be neglected in our theoretical investigation. This means that our results should apply to the temperature region 
for which the correlation length is smaller than the average distance between two defects in a real spin chain.         
Under these hypotheses, some DWs will be always present in the system at finite temperature. 
A simple random-walk argument relates the relaxation time, $\tau$, to the correlation length~\cite{Tobochnik}:  
within a time $\tau$ a domain wall performs a random walk over a distance $\xi$. 
The characterization of SCMs basically consists in measuring both these quantities ($\xi$ and $\tau$) as function of temperature. 
In ideal cases, the observed behavior is then reproduced by fitting the parameters of an appropriate spin model to the experimental data. 

Following the experimental procedure, we studied the temperature dependence of the correlation length and the relaxation time in a representative model 
for classical spin chains with uniaxial anisotropy. $\xi$ has been computed with the transfer-matrix technique~\cite{Fisher63AJP,Blume75PRB,Vindigni06APA}. 
The relaxation of the magnetization and DW diffusion have been studied by using time-quantified Monte 
Carlo (TQMC)~\cite{Nowak00PRL,Cheng06PRL,Billoni07JMMM}. The latter is a recently developed algorithm which, 
in classical spin systems, simulates a dynamics equivalent to the stochastic Landau-Lifshitz-Gilbert equation. 
TQMC has been previously employed to model classical spin chains in the high damping limit~\cite{Hinzke00PRB,Hinzke01Proc}, but the 
most recent developments allow using it for low-damping calculations~\cite{Cheng06PRL}. 

Most of SCMs show a large single-site magnetic anisotropy. 
The reference model is thus represented by the 1d Ising model or -- more specifically -- its kinetic version proposed by Glauber~\cite{Glauber63JMathP}. 
In order to better capture the physics of experimental systems,  
the Glauber model have been extended to take into account finite-size effects~\cite{Barma_J_Stat_Phys_80,Luscombe_PRE_96,daSilva_PRE_95,PRL_Bogani,Coulon04PRB,Vindigni_JMMM_04},  
ferrimagnetism~\cite{Pini-Rettori_PRB_07}, the effect of a strong external magnetic field~\cite{Coulon_PRB_07} and the reciprocal non-collinearity of local anisotropy axes~\cite{Pini-Vindigni_JPCM_09,Caneschi02EPL,Bernot_PRB_09}.  
Here we report about a linear ferromagnetic spin chain and we address the question of how physics changes when the single-site anisotropy is progressively reduced. 
We found that the Glauber scenario needs to be revisited for SCMs which do not possess 
a large single-site anisotropy so that DWs acquire a finite thickness (more than one lattice unit). 
Our theoretical predictions are in agreement with the few available experimental results on molecular spin chains with low single-site anisotropy~\cite{Balanda,Miyasaka06CEJ}. 
Note that in metallic nonowires (Co, Ni, Fe, Permalloy) DWs always extend over several lattice units. 
Therefore, for  materials traditionally used in magneto-storage manufactory     
the Glauber's picture of magnetization reversal and relaxation is not expected to hold exactly, nor the correlation length is expected to show an Ising-like behavior    
over a wide temperature range.    

In section~\ref{sec_II}, we present the model and define the physical questions that we want to address. 
In section \ref{sec_III}, we study the temperature dependence of the correlation length by means of the transfer-matrix technique and Polyakov renormalization. 
In section \ref{sec_IV}, we introduce the time-quantified Monte Carlo method and study the temperature dependence of the relaxation time in the broad-wall regime. 
In section \ref{sec_V}, we use TQMC to study temperature-induced diffusion of both sharp and broad DWs. 
In section \ref{sec_VI}, we provide some phenomenological arguments which support our findings and discuss how they compare with
experiments. 
In the conclusions, we summarize our main results and highlight their relevance for further work which could possibly include an electric 
current driving DW motion.    
\section{The system\label{sec_II}} 
%%% FIGURA 1 A %%%%%%%%%%%%%%%%%%%%%%%%%%%%%%%%%%%%%%%%%%%%%%%%%%%%%%%%%%%%%%%%%%%%%%%%%%%%%
\begin{figure}[b]%[ht]
\includegraphics*[width=8.5cm,angle=0]{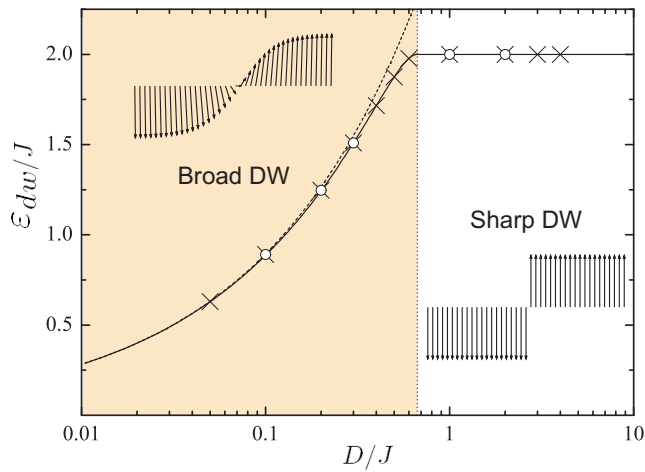}
\caption{\label{fig1_intro} 
Domain-wall energy (in $J$ units) as a function of the ratio $D/J$, adapted from  Ref.~\onlinecite{Vindigni08ICA}: 
discrete-lattice calculation $\varepsilon_{dw}$ (solid line), continuum-limit solution $2\sqrt{2DJ}$ (dashed line). Crosses and circles correspond to 
the ratios used in transfer-matrix and time-quantified Monte-Carlo calculations respectively. The vertical line 
indicates the value ($D/J \!=\!2/3$) at which the transition from broad- to  sharp-wall regime occurs.  
Insets:  sketch of the spatial dependence of $S_i^z$ as a function of $i$ in the broad-wall (up left) and in the sharp-wall (down right) regime. }
\end{figure}
%%%%%%%%%%%%%%%%%%%%%%%%%%%%%%%%%%%%%%%%%%%%%%%%%%%%%%%%%%%%%%%%%
As a reference model for SCMs, we consider the following classical Heisenberg Hamiltonian:
%
%%%  EQUATION 1: Hamiltonian %%%%%%%%%%%%%%%%%%%%%%%%%%%%%%%%%%%%%%%%%%%%%%%%%%%%%%%%%%%%%%%%%%%%%%%%%%%%%%
\begin{equation}\label{eq1}
H=- \sum_{i=1}^N \left[ J \vec{S}_{i} \cdot \vec{S}_{i+1} + D (S_{i}^z)^2 + \vec{H} \cdot \vec{S}_i\right]
\end{equation}
%%%%%%%%%%%%%%%%%%%%%%%%%%%%%%%%%%%%%%%%%%%%%%%%%%%%%%%%%%%%%%%%%%%%%%%%%%%%%
%
\noindent where $D$ represents the anisotropy energy, $J$ the exchange coupling and $\vec{H}$  
an external applied field. Each spin variable $\vec{S}_i$ is a three-component unit vector associated with 
the $i$--th node of the lattice. In this paper either periodic % boundary conditions $i\!+\!N\!=\!i$ or 
or open boundary conditions will be considered depending on the calculation. $D$ and $J$ are assumed positive so that Hamiltonian~\eqref{eq1} 
describes a spin chain with uniaxial anisotropy pointing in the $z$ direction.  
This model and variations of it have been investigated extensively from the theoretical point of view~\cite{Krumhansl75PRB, Nakamura78JPCSSP,Fogedby84JPCSSP,Hinzke00PRB,Seiden}. 
Moreover, Hamiltonian~\eqref{eq1} has been employed to reproduce the experimental behavior of some SCMs~\cite{Vindigni06APA,Coulon04PRB}.   
Many physical properties are related to 
the energy increase due to the creation of a domain wall (DW) in one 
of the two ground-state, uniform configurations with $S_i^z\!=\!\pm 1$ ($\forall i$). 
In Fig.~\ref{fig1_intro} we plot this energy 
obtained from a discrete-lattice calculation  $\varepsilon_{dw}$  (solid line) 
for different values of the ratio $D/J$. 
For $D/J\!\le\!2/3$, the minimal DW energy is realized by a spin profile in which 
several spins are not aligned along the easy axis, $z$ (see the sketch in the inset up on the left). 
In this case, DWs spread over more than one lattice spacing: {\it broad } DWs. 
For $D/J\!>\!2/3$, the minimum DW energy is realized if the 
transition between $S_i^z\!=\!+ 1$ to $S_i^z\!=\!-1$ occurs within one lattice
spacing; in this way all the spins are aligned along the easy axis:
{\it sharp } DWs (see the sketch in the inset down on the right). 
The transition between broad- and sharp-wall regime is highlighted by a 
singularity in the log-linear plot, which  evidences a different functional dependence 
of $\varepsilon_{dw}$ on the parameters $D$ and $J$ for anisotropy-to-exchange ratios smaller or larger than 2/3. 
For $D/J\! \ll \! 1$, the analytic expression $2\sqrt{2DJ}$ can be obtained 
by minimizing the DW energy in the continuum-limit approximation~\cite{Jongh74AP} (see Eq.~\eqref{Heisenberg_Ham_cont}). 
This function is plotted in Fig.~\ref{fig1_intro} as a dashed line. 
Approaching the transition ratio, $D/J\!=\!2/3 $,  from below the continuum-limit prediction starts to deviate from the discrete-lattice calculation,  $\varepsilon_{dw}$ (solid line). 
This is reasonable since the continuum approximation is more accurate for smaller anisotropy-to-exchange ratios, namely the broader DWs are. 
For $D/J\!>\!2/3$, the DW energy takes the constant value $\varepsilon_{dw}\!=\!2J$. 
The transition region where none of the two analytic expressions  
holds exactly is narrow, meaning that the  function 
\begin{equation}
\label{definition_E_w}
\varepsilon_{w} = 
\begin{cases}
\,2\sqrt{2DJ} \quad &\text{for}  \quad  D/J \le 2/3  \\
\, 2 J  \quad &\text{for}  \quad  D/J > 2/3 
\end{cases}
\end{equation}
describes the DW energy accurately for most values of $D$ and $J$.  
In Fig.~\ref{fig1_intro}, symbols represent the DW energy  $\varepsilon_{dw}$ for  
the values of $D/J$ at which static  (circles) and dynamic (crosses) calculations have been performed.   \\
The main question we want to address is the following: are SCMs ruled by different laws depending on whether DWs are 
sharp or broad? Our results show that physics is significantly different in these two limits. This fact will eventually affect 
the \textit{experimental} characterization of SCMs which can be  modeled by Hamiltonian~\eqref{eq1} 
(or variations of it~\cite{Balanda,Coulon04PRB,Clerac02JACS,Miyasaka06CEJ,Vindigni06APA,Caneschi02EPL,Bernot_PRB_09}).  \\
Two key quantities characterizing a specific SCM are the correlation length, $\xi$, and the relaxation time, $\tau$. 
For our study, the most relevant correlations are those relating to spin projections along the easy axis. 
Therefore, the correlation length shall be defined from pair-spin correlations along $z$ as:
%
%%% EQUATION 2: Correlation length %%%%%%%%%%%%%%%%%%%%%%%%%%%%%%%%%%%%%%%%%%%%%%%%%%%%%%%%%%%%%%%%%%%%%%
\begin{equation}\label{xi_def}
\langle S_{i+r}^z S_{i}^z\rangle = \langle (S_{i}^z)^2\rangle e^{-\frac{r}{\xi}}, 
\end{equation}
%%%%%%%%%%%%%%%%%%%%%%%%%%%%%%%%%%%%%%%%%%%%%%%%%%%%%%%%%%%%%%%%%%%%%%%%
%
\noindent (where $ \langle \dots \rangle $ stands for thermal average). The static susceptibility along $z$ scales with the correlation length as follows:
%
%%% EQUATION 3: Static susceptibility %%%%%%%%%%%%%%%%%%%%%%%%%%%%%%%%%%%%%%%%%%%%%%%%%%%%%%%%%%%%%%%%%%%
\begin{equation}\label{xi_chi_stat}
\chi_z(T)\sim \frac{\xi}{T}\,. 
\end{equation}
%%%%%%%%%%%%%%%%%%%%%%%%%%%%%%%%%%%%%%%%%%%%%%%%%%%%%%%%%%%%%%%%%%%%%%%%%%%%%%%%%%%%%%%%%%%%%%%%%%%%%%%%%
%
%%% EQUATION 4: Dynamic susceptibility %%%%%%%%%%%%%%%%%%%%%%%%%%%%%%%%%%%%%%%%%%%%%%%%%%%%%%%%%%%%%%%%%
The relaxation time can be obtained from the dynamic susceptibility,
\begin{equation}\label{chi_ac}
\chi(\omega, \,T) = \frac{\chi(T)}{1-i\omega\tau} \,,
\end{equation}
%%%%%%%%%%%%%%%%%%%%%%%%%%%%%%%%%%%%%%%%%%%%%%%%%%%%%%%%%%%%%%%%%%%%%%%%%%%%%%%%%%%%%%%%%%%%%%%%%%%%%%%
%
\noindent where $\omega$ is the frequency of the oscillating applied field and $\chi(T)$ is the static susceptibility.  Both the real and the imaginary part 
of $\chi(\omega, \,T)$ display a maximum for $\omega\tau\!=\!1$. 
In the following we will use an alternative definition of $\tau$ in terms of the relaxation of the magnetization.  
Eqs.~\eqref{xi_chi_stat} and~\eqref{chi_ac} relate $\xi$ and $\tau$ to measurable quantities, the static and dynamic 
susceptibility. 
Based on a random-walk argument~\cite{Tobochnik}, the relation between the correlation 
length and the relaxation time
%%% EQUATION 5: Correlation length, relaxation time nexus %%%%%%%%%%%%%%%%%%%%%%%%%%%%%%%%%%%%%%%%%%%%%%
\begin{equation}\label{xi-tau}
\xi^2\simeq 2D_s\tau  
\end{equation}
%%%%%%%%%%%%%%%%%%%%%%%%%%%%%%%%%%%%%%%%%%%%%%%%%
\noindent 
is usually assumed~\cite{Miyasaka_review,Coulon06Springer,Bogani_JMC_08} for such temperatures 
that $\xi \! < \! N$ (bulk regime).
$D_s$ is the DW diffusion coefficient but it can also be interpreted as the attempt frequency for a 
single-spin flip. The temperature dependence of $D_s$ will be discussed in more details in Sect.~\ref{sec_V}.    
Within the kinetic Ising model, it is possible to deduce Eq.~\eqref{xi-tau} analytically~\cite{Glauber63JMathP,Tobochnik}. 
The basic experimental characterization of a SCM essentially reduces to determining the temperature dependence of the 
three quantities  involved in Eq.~\eqref{xi-tau}: $\tau$, $\xi$ and $D_s$. The scenario is well-established in the 
sharp-wall regime, in which one  expects that all these quantities obey a thermally activated mechanism: 
%
%%% EQUATION 6 Thermally activated mechanism  %%%%%%%%%%%%%%%%%%%%%%%%%%%%%%%%%%%%%%%%%%%%%%%%%%%%%%%
\begin{eqnarray}\label{activation_xi-tau-D}
\xi\sim e^{\frac{\Delta_\xi}{T}}\quad\quad
\tau\sim e^{\frac{\Delta_\tau}{T}}\quad\quad
D_s\sim e^{-\frac{\Delta_A}{T}} \,.
\end{eqnarray}
%%%%%%%%%%%%%%%%%%%%%%%%%%%%%%%%%%%%%%%%%%%%%%%%%%%%%%%%
%
\noindent $k_B \!=\!1$ will be assumed throughout the manuscript. 
Eq.~\eqref{xi-tau} relates $\xi$ and $\tau$ -- two experimentally accessible quantities -- 
with each other so that the following relation between energy scales holds:
\begin{equation}\label{Delta_tau}
\Delta_\tau = 2\Delta_\xi+ \Delta_A\,. 
\end{equation}
The above relation has been confirmed by several experimental works on sharp-wall SCMs~\cite{Miyasaka_review,Coulon06Springer,Bogani_JMC_08} using the reasonable 
assumption $\Delta_A\simeq D$ proposed in Ref.~\onlinecite{Coulon04PRB,Kirschner97}. In Sect.~\ref{sec_V}
we will see that our numerical results confirm the validity of Eq.~\eqref{Delta_tau} for sharp DWs. On the contrary, the scenario turns out to 
be significantly different in the broad-wall regime. In fact, for $D/J\! < \!2/3 $, we found that: 
\begin{itemize} 
\item $\Delta_\xi$ is not constant but is reduced by more than 30\% of the value that it takes at low $T$ with 
increasing temperature.
\item $D_s$ does not follow a thermally activated mechanism as suggested by Eq.~\eqref{activation_xi-tau-D}.  
\end{itemize}   
These two important findings indeed affect the experimental characterization of a SCM for which 
$D/J \!<\!2/3 $.   
\section{Static properties\label{sec_III}} 
\subsection{Transfer matrix calculations}
\begin{figure}%[ht]
\includegraphics*[width=8cm,angle=0]{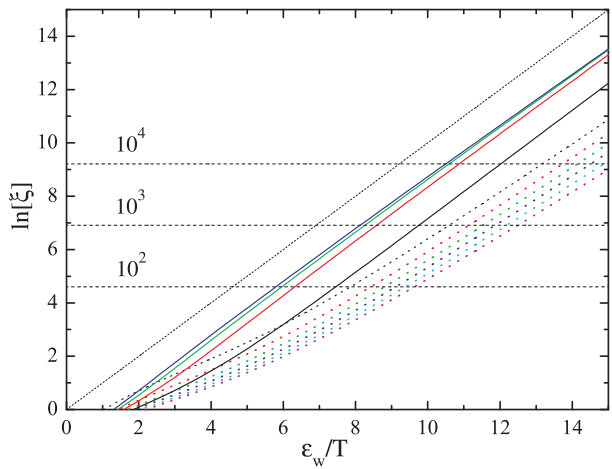}
\caption{\label{fig1} Color online. Logarithm of the correlation length obtained by TM calculations as function of the
inverse temperature in units of the DW energy $ \varepsilon_w$ 
(defined  in Eq.~\eqref{definition_E_w}) for $J=1$ and different values of $D$: 
Dotted lines, \textit{broad} domain walls, 
$D\!=\!$ 0.1 (black), 0.2 (red), 0.3 (green), 0.4 (blue), 0.5 (cyan), 0.6 (magenta).   
Solid lines, \textit{sharp} domain walls,  
$D\!=\!$ 1 (black), 2 (red), 3 (green), 4 (blue).  
Dashed horizontal lines give the reference for typical lengths of realistic spin chains. } 
\end{figure}

The static properties of a classical spin chain can be computed efficiently with the transfer-matrix (TM) method~\cite{Tannous,Blume75PRB}. 
Here we use this technique to compute the thermodynamic properties of an infinite chain ($N\rightarrow\infty$) but finite systems could also be considered~\cite{Vindigni06APA}. 
The essential ideas of the TM approach are recalled in Appendix~\ref{TM_Appendix}.  In particular, we used this method 
to compute the correlation length. The dependence on $r$ can be eliminated from Eq.~\eqref{xi_def} by summing over all
the lattice separations 
\begin{equation}
\sum_{r\ge 0} \langle S_{i+r}^z S_{i}^z\rangle = \langle (S_{i}^z)^2\rangle\sum_{r\ge 0} e^{-\frac{r}{\xi}}
=\frac{\langle (S_{i}^z)^2\rangle}{1-e^{-1/\xi}}\, . 
\end{equation}
In practice, we evaluated the summation numerically and inverted the previous formula as follows
\begin{equation}
\xi=-\left[\ln\left(1-\frac{\sum_{r\ge 0} \langle S_{i+r}^z S_{i}^z\rangle}{\langle (S_{i}^z)^2\rangle}\right)\right]^{-1}\,.
\label{xi_def_TM}
\end{equation}
Fig.~\ref{fig1} shows the logarithm of the correlation length {\it vs.} the energy of the DW 
divided by the temperature as obtained by TM calculations. Different curves correspond to different values of  $D/J$. 
Since we set $J\!=\!1$, the cases with $D\!=\! 1 - 4$ fall in the sharp-wall regime,  while the cases with 
$D\!=\! 0.1 - 0.6$ correspond to the broad-wall regime (see Fig.~\ref{fig1_intro}). We will discuss the behavior of $\xi$ in these two regimes separately. \\
%%
%
%%% FIGURA 1 
%%%%%%%%%%%%%%%%%%%%%%%%%%%%%%%%%%%%%%%%%%%%%%%%%%%%%%%%%%%%%%%%%%%%%%%%%%
%%%%%%%%%%%%%%%%%%%%%%%%%%%%%%%%%%%%%%%%%%%%%%%%%%%%%%%%%%%%%%%%%%%%%%%
%
\textit{Sharp-wall regime} -- 
In this regime, the DW energy amounts to $\varepsilon_w\!=\!2J$. 
This value has been used to plot the logarithm of the correlation length as a function of $\varepsilon_w/T$ for  $D\!=\! 1 - 4$ in Fig.~\ref{fig1}. 
At low temperatures, all the solid curves have the same slope, equal to one, revealing that $\ln(\xi)\! \sim\!  \varepsilon_w/T$.  
A reference line with slope one is plotted with short dashes. For $D\!=\!1$,  some deviations from this straight line occur at high temperatures.  
Later on, we will show that in the broad-wall regime the interplay between spin waves and DWs leads to an effective decrease of $\Delta_\xi$ with increasing $T$.   
It is reasonable to think that a similar phenomenon may take place in the sharp-wall regime as well, when the transition ratio $D/J\!=\!2/3$ is approached from above.  
The horizontal dotted lines indicate the values 
$\xi\!=\!10^2,10^3,10^4$. Due to lattice dislocations or impurities (in molecular compounds) or intrinsic problems 
in the deposition procedure (in mono-atomic nanowires~\cite{Nature_Gambardella}), the average length of 
spin chains is typically of $10^2 - 10^4$ magnetic centers in real SCMs. 
The actual length of a spin chain sets an upper bound to the 
low-temperature divergence of $\xi$. Such an upper bound can be reduced by introducing additional non-magnetic 
impurities~\cite{PRL_Bogani}. Most of the SCMs reported in the literature fall in the sharp-wall regime~\cite{Miyasaka_review,Coulon06Springer,Bogani_JMC_08}. 
For them, in the temperature range where the correlation-length divergence is relevant ($\xi \!\gg \!1$), $\Delta_\xi$ 
can be assumed equal to $2J$, independently of $T$.  \\
%%
%%%%%%%%%%%%%%%%%%%%%%%%%%%%%%%%%%
%%
\textit{Broad-wall regime} -- 
For $D/J\!<\!2/3$, the DW energy is described well by the function $\varepsilon_w\!=\!2\sqrt{2DJ}$ 
(see Eq.~\eqref{definition_E_w}).  As already pointed out when discussing Fig.~\ref{fig1_intro},
this expression obtained in the continuum limit 
deviates from the discrete-lattice calculation 
in the vicinity of the transition region from broad to sharp DWs.
In order to make the comparison with experiments easier,    
in the main frame of Fig.~\ref{fig1} the correlation length is plotted {\it vs.} $\varepsilon_w/T$, with $\varepsilon_w\!=\!2\sqrt{2DJ}$ for all values of $D$ falling in the broad-wall regime
(instead of using $\varepsilon_{dw}$, {\it i.e.}, value obtained from the discrete-lattice calculation). 
At low temperature the slope of all the dotted lines is about 0.9, which suggests an effective $\Delta_\xi=1.8\sqrt{2DJ}$. 
This ten per cent of reduction with respect to the DW energy can be accounted for using the low-temperature expansion of the 
correlation  given in Ref.~\onlinecite{Fogedby84JPCSSP}: $\xi\sim \left(T/\varepsilon_w\right)\exp\left( \varepsilon_w/T\right)$. 
More strikingly, all the dotted curves bend when the temperature increases till their slope becomes roughly 0.6  at high temperature. 
A tentative fit of the temperature dependence of $\xi$ using the corresponding formula in Eq.~\eqref{activation_xi-tau-D} would give a 
value of the activation energy  $1.2\sqrt{2DJ} \!< \!\Delta_\xi \!< \! 1.8\sqrt{2DJ}$ depending on the temperature range in which 
the fitting has been performed. This is nothing but the standard procedure followed in the experimental characterization of 
SCMs~\cite{Miyasaka_review,Coulon06Springer,Bogani_JMC_08}. 
Normally, the extracted energy is then related to the barrier observed in the relaxation time by means of  
Eq.~\eqref{Delta_tau}. It is worth remarking that the temperature ranges where $\Delta_\xi$ and $\Delta_\tau$ are extracted usually 
do not overlap. In fact, to measure $\xi$ the characteristic time scale of the experiment has to be much longer than the relaxation time; 
while for measuring  $\tau$ the characteristic time scale of the experiment has to be comparable to the relaxation time itself, or shorter.  
Thus, the \textit{effective} variation of $\Delta_\xi$ with $T$ -- which may reach 30\% of its value -- has to be considered when one tries to relate this quantity to $\Delta_\tau$ by means of Eq.~\eqref{Delta_tau}.  
In particular, we remark again that $\Delta_\xi$ can only be accessed at relatively high temperature because 
finite-size effects or three-dimensional inter-chain interactions prevent the  correlation length
from diverging indefinitely~\cite{Miyasaka06CEJ,Balanda}.  \\
%%%
As an example of how this first theoretical result may be used to better characterize real SCMs, we refer to the 
system reported in  Ref.~\onlinecite{Balanda}. That spin chain is better described by the Seiden model~\cite{Seiden} with anisotropy 
rather than Hamiltonian~\eqref{eq1}.  
Apart from a multiplicative factor two in front of the exchange-energy term, the Hamiltonian 
of the Seiden model and the one we study here take the same form in the continuum limit~\cite{Coulon_private} (see Eq.~\eqref{Heisenberg_Ham_cont}).  
Yet, the experimental system shows $\Delta_\xi^{exp}\simeq60$ K and $J\!=\!108.5$ K 
(adapting the fitted value to our model). Thus
\begin{enumerate} 
\item assuming that the fit of $\Delta_\xi^{exp}$ has been performed in the region where 
$\Delta_\xi\!=\!0.6 \varepsilon_w$, one would obtain $\varepsilon_w\!\simeq\!100$ K and, accordingly,  $D\!=\!11.5$ K;  
\item assuming that in the fitted region $\Delta_\xi\!=\!0.9 \varepsilon_w$, one would get $D\!=\!5.1$ K. 
\end{enumerate}  
The first estimate of $D$ falls in the suggested range $D\!=\! 8.2 - 17.2 $ K 
(estimated by EPR measurements on equivalent isolated magnetic units~\cite{Balanda}) while the second estimate is clearly wrong. 
The experimental correlation length displays a clear exponential divergence in the range 25 K $<\!T\!<$ 50 K, which corresponds to 
$2\!<\!\varepsilon_w/T\!<\!4$ (if $\varepsilon_w\!=\!100$ K is assumed).  
Indeed, these values of the reduced variable $\varepsilon_w/T$ relate to the high-temperature region where 
$\Delta_\xi\!\simeq\!0.6 \varepsilon_w$ for $D/J\simeq 0.1$ (see Fig.~\ref{fig1}).

In the following we will give a justification of the important variation of $\Delta_\xi$ with $T$ observed in the broad-wall regime in terms of 
Polyakov renormalization~\cite{Polyakov}. 
\subsection{Polyakov renormalization}  
We focus now in the broad-wall regime in which $\Delta_\xi$   decreases with increasing temperature. 
A deeper insight in such a phenomenon can be achieved by considering the effect of spin-wave renormalization 
on the coupling constants $D$ and $J$. To this aim, we work with the continuum version of the Hamiltonian 
in Eq.~\eqref{eq1}: 
\begin{equation} 
\label{Heisenberg_Ham_cont}
\mathcal{H}=-JN + \int \left[ \frac{J}{2}|\partial _x \vec{S}|^2
-D \left(S^z(x)\right)^2\right] dx\,,
\end{equation}
where a unitary lattice spacing has been assumed. 
Following Polyakov~\cite{Polyakov,Politi_EPL_94}, we represent $\vec{S}(x)$ as a superposition of 
a field fluctuating over \textit{short} spatial scales, $\vec{\phi}(x)$, and a field varying smoothly and over \textit{large} spatial scales, $\vec{n}(x)$. 
More explicitly, we write 
\begin{equation} 
\vec{S}(x) = \vec{n}(x)\sqrt{1-\phi^2(x)} + \vec{\phi}(x)\,.
\end{equation} 
Requiring $|\vec{S}(x)|\!=\!1$ and $|\vec{n}(x)|\!=\!1$, one necessarily has $\vec{n}(x)\!\cdot\!\vec{\phi}(x)\!=\!0$; thus 
$\vec{\phi}(x)$ can be expressed on a basis orthonormal to $\vec{n}(x)$: 
\begin{equation} 
 \vec{\phi}(x)= \sum_a \phi_a \vec{e}_a
\quad\quad\text{with  } a=1,2
\end{equation}  
with $|\vec{e}_a(x)|\!=\!1$. 
The terms appearing in Hamiltonian~\eqref{Heisenberg_Ham_cont} are affected by the 
averaging procedure over the $\vec{\phi}(x)$ field as follows: 
\begin{equation} 
\label{Polyakov_eqs}
\begin{cases}
\langle|\partial _x \vec{S}|^2\rangle\!&=\left[1-\langle \phi_a^2 \rangle \right]  \left(\partial _x\vec{n}\right)^2 
+ \sum_{a} \langle\left( \partial_x \phi_a \right)^2 \rangle \\
\langle\left(S^z\right)^2 \rangle \!&= \left[1-3\langle \phi_a^2\rangle\right] \left( n^z\right)^2 + \langle \phi_a^2\rangle\,.
\end{cases}
\end{equation}
The latter equations represent a well-known result~\cite{Polyakov,Politi_EPL_94}, however we will recall their derivation in Appendix~\ref{Polyakov} 
for convenience. \\
%
%%%%%%%%%%%%%%%%%%%%%%%%%%%%%%%%%%%%%%%%%%%%%%%%%%%%%%%%%%%%%%%%%%%%%%%%%%%%%%%
%
In the isotropic Heisenberg chain ($D\!=\!0)$ and in the easy-plane case ($D\!<\!0)$ 
excited spin waves suffice to destroy the long-range order present in the ground state at finite temperatures
(because of the existence of a Goldstone mode).   
When $D\!>\!0$, the spectrum of spin-wave excitations acquires a gap so that the 
long-range order is rather destroyed by DW proliferation at finite temperatures.
To fix the ideas, one can think the field $\vec{n}(x)$ to be associated with localized excitations (DWs) while  
$\vec{\phi}(x)$ being associated with spin waves. 
Within a distance separating two successive DWs one can assume 
$\left(n^z\right)^2 \simeq 1 $ and $ |\partial _x \vec{n}|^2\simeq 0 $.
Therefore, up to quadratic terms, the $\vec{\phi}$-field Hamiltonian reads: 
\begin{equation} 
\label{SW_Ham}
\mathcal{H}_{\phi} = \frac{N}{\Lambda}\sum_{a}  \int_{-\Lambda/2}^{\Lambda/2}  \left[ \frac{J}{2} \left( \partial_x \phi_a \right)^2 + D \phi_a^2 \right] dx \, ,
\end{equation}
$\Lambda$ being the average distance between two successive DWs
(see Appendix~\ref{Polyakov} for the derivation of Eq.~\eqref{SW_Ham}).  
%%%%%%%%%%%%%%%%%%%%%
Following the typical procedure of the renormalization group (RG),  we express the Hamiltonian \eqref{SW_Ham} in the 
Fourier space 
\begin{equation} 
\label{SW_Ham_Fourier}
\mathcal{H}_{\phi} = \frac{N}{\Lambda}\sum_{a}  \frac{1}{2\pi}
\int \left( \frac{J}{2} q^2 + D  \right) |\tilde{\phi}_a(q)|^2  dq \, ,
\end{equation}
and apply equipartition 
\begin{equation} 
\label{Equipartition}
\langle |\tilde{\phi}_a(q)|^2 \rangle = \frac{T}{Jq^2+2D}\,.
\end{equation}
The integration of the \textit{fast}-fluctuating excitations in the range $\frac{\pi}{L'}  \leq q \leq \frac{\pi}{L}$, with 
$L'=L+dL$, yields 
\begin{equation} 
\label{differential_phi}
%\begin{split} 
\langle \phi_a^2 \rangle = \frac{T}{2\pi} \int^{\pi/L}_{\pi/L'} \frac{dq}{Jq^2+2D}= 
 \frac{T}{4 D}\frac{dL}{L^2+\left(\frac{\pi w}{L}\right)^2}\,,
%\end{split} 
\end{equation} 
where $w^2\!=\! J/2D$. 
The prefactors of $\left(\partial _x\vec{n}\right)^2$ and $\left( n^z\right)^2$  in Eqs.~\eqref{Polyakov_eqs}
together with Eq.~\eqref{differential_phi} define the RG equations: 
\begin{equation}
\label{RG_equations} 
\begin{cases} 
&J(L+dL) = J(L) \left[1-\langle \phi_a^2 \rangle \right]   \\
&D(L+dL) = D(L) \left[1-3\langle \phi_a^2\rangle\right]  \,.
\end{cases} 
\end{equation}
As only short-range order is present in a 1d system, it is reasonable to integrate Eqs.~\eqref{RG_equations} 
only up to the average distance separating two successive DWs,  $\Lambda$.  
The inverse of $\Lambda$ can be identified with the average density of DWs. A low-temperature 
estimate of the DW density is given in Ref.~\onlinecite{Fogedby84JPCSSP}:
\begin{equation}
\label{Fogedby_eq} 
\frac{1}{\Lambda} = \frac{1}{w}\frac{4 \varepsilon_w }{T}\exp\left(-\frac{\varepsilon_w}{T}\right)
\end{equation}
where $\varepsilon_w\!=\!2\sqrt{2JD}$ again. 
Note that in Ref.~\onlinecite{Fogedby84JPCSSP} the DW density is given in terms of 
\textit{bare} constants $J$ and $D$. 
Our goal is to see how $\Lambda$ has to be corrected at finite temperature to account for spin-wave renormalization. 
This has been done with the following procedure:    
\begin{enumerate}
\item fix the temperature $T$
\item compute numerically the renormalized constants at an infinitesimal $dL$ starting from the \textit{bare} 
values of  $J$ and $D$
\item compute $\Lambda(dL)$ from Eq.~\eqref{Fogedby_eq} using the renormalized constants $J(dL)$ and $D(dL)$
\item iterate the procedure till $L+dL > \Lambda(L)$ 
\item at this point we defined $\tilde{\Lambda}(T)=\Lambda(L)$ and analogously the renormalized 
 constants:  $\tilde{D}(T)$, $\tilde{J}(T)$, $\tilde{\varepsilon}_w(T)$ and $\tilde{w}(T)$. 
\end{enumerate}
Apart from a multiplicative constant, the final value of $\tilde{\Lambda}(T)$ 
and the correlation length, $\xi$, are supposed to depend on temperature likewise.   \\
Before proceeding with the analysis of the RG equations, we specify the convention on the notation we use.  
In this section, we introduced a dependence on $L$ in the quantities $D(L)$, $J(L)$, $\varepsilon_w(L)$ and $w(L)$ in order to renormalize them.   
This dependence is somewhat technical. The relevant values of those renormalized quantities are 
$\tilde{D}(T)$, $\tilde{J}(T)$, $\tilde{\varepsilon}_w(T)$ and $\tilde{w}(T)$, namely the values assumed 
when the renormalization stops. These last values only depend on temperature, not on $L$ anymore. 
In the other sections, when not specified differently, we always refer to the \textit{bare} 
constants; these equal the values of the corresponding renormalized quantities at $T\!=\!0$ and those of the 
``technical'' $D(L)$, $J(L)$, $\varepsilon_w(L)$ and $w(L)$ for $L\!=\!0$. 
%  
%%%%%%%%%%%%%%%%%%%%%%%%%%%%%%%%%%%%%%
\begin{figure}
\begin{center}               
\includegraphics[width=8.5cm]{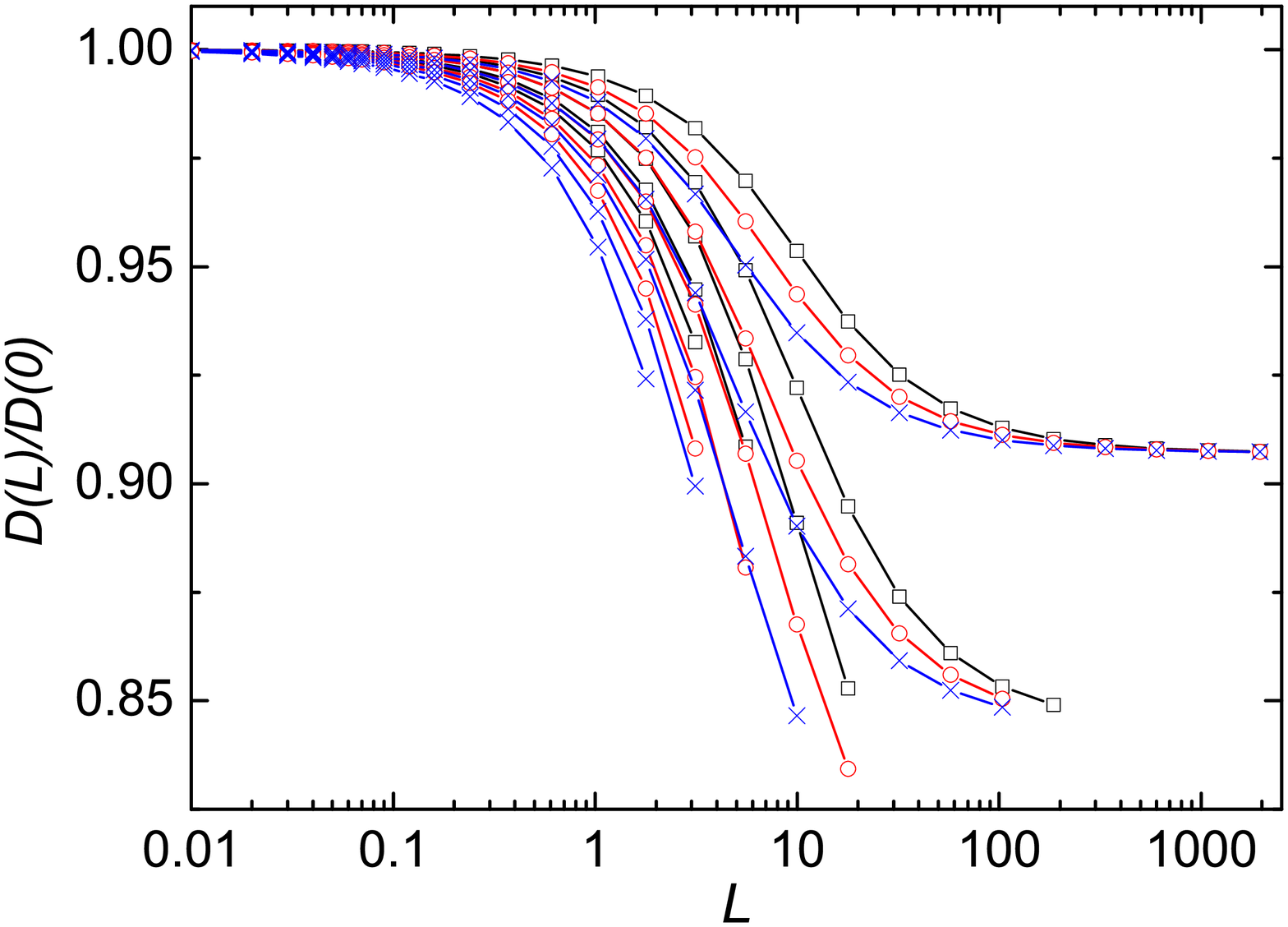}
\end{center}
\caption{Color online. The ratio between renormalized constant $D(L)$ and the bare constant $D(0)$ 
is plotted as a function of some selected intermediate values of $L$,
for different values of  $D(0)=0.05$ (black squares) ,0.1 (red circles), 0.2 (blue crosses) 
and different temperatures $\varepsilon_w(0)/T=$ 0.0625, 0.106, 0.149, 0.193, 0.236.   
The renormalization flux stops at  $L=\tilde{\Lambda}(T)$.
\label{Fig_11}}
\end{figure}
%%%%%%%%%%%%%%%%%%%%%%%%%%%%%%%%%
%%%%%%%%%%%%%%%%%%%%%%%%%%%%%%%
\subsection{Symmetry properties of \\the renormalization flux\label{Symm_Ren_flux}} 
%%%%%%%%%%%%%%%%%%%%%%%%%%%%%%%%%%%%%%
%%%%%%%%%%%%%%%%%%%%%%%%%%%%%%%
\begin{figure}
\begin{center}               
\includegraphics[width=8.5cm]{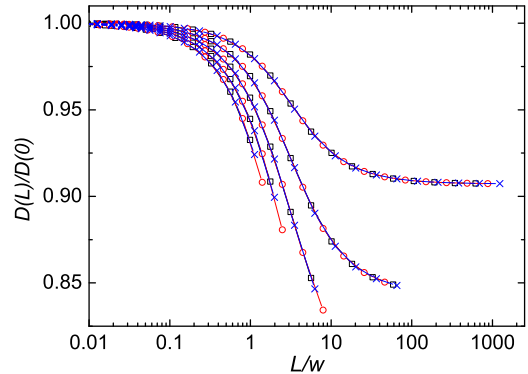}
\end{center}
\caption{Color online. Same data as in Fig.~\ref{Fig_11} but plotted \textit{vs.} $L/w(0)$. \label{Fig_12}}
\end{figure}
%%%%%%%%%%%%%%%%%%%%%%%%%%%%%%%%%
%%%%%%%%%%%%%%%%%%%%%%%%%%%%%%%
By noting that $4 w D= \varepsilon_w$ one can rewrite the average of the $\vec{\phi}$-field components as: 
\begin{equation} 
\label{differential_phi_2}
\langle \phi_a^2 \rangle = \frac{T}{\varepsilon_w}\frac{w}{L} \frac{1}{1+\left(\frac{\pi w}{L}\right)^2} \frac{dL}{L} \,.
\end{equation} 
From the above expression, one can immediately see that  $\langle \phi_a^2 \rangle $ depends on 
$L/w(0)$ and $\varepsilon_w/T$ only. 
This symmetry property is directly transferred to the 
RG equations. This can be made more transparent by rewriting Eqs.~\eqref{RG_equations} as 
\begin{equation}
\label{RG_equations_rescaled} 
\begin{cases} 
&\frac{dJ}{J} = -  \frac{T}{\varepsilon_w}\frac{w}{L} \frac{1}{1+\left(\frac{\pi w}{L}\right)^2} \frac{dL}{L}  \\
&\frac{dD}{D} =-3  \frac{T}{\varepsilon_w}\frac{w}{L} \frac{1}{1+\left(\frac{\pi w}{L}\right)^2} \frac{dL}{L} \,. 
\end{cases} 
\end{equation}
Note that the renormalized constants will, however, depend on the values they take at zero temperature $\tilde J(0)$ and $\tilde D(0)$, {\it i.e.} the bare 
constants. Without loss of generality, we can set $J=1$ (as a unit for the energies and temperature) and 
focus on the renormalization of $D$ with increasing $L$. In order to compare the renormalization of the anisotropy energy with $L$ for different
values of the bare $D(0)$, we plot the ratio $D(L)/D(0)$ (Fig.~\ref{Fig_11}). 
The different temperatures have been chosen so that the ratios $\varepsilon_w(0)/T$ are the same for different initial anisotropy values. 
Fig.~\ref{Fig_12} shows how all the curves corresponding to the same  $\varepsilon_w(0)/T$ ratio but different $D(0)$ collapse onto a single one when plotted as a function of $L/w(0)$ (instead 
of $L$). 
Such a data collapsing evidences that due to the symmetry of Eqs.~\eqref{RG_equations_rescaled} the renormalization flux, {\it i.e.} the relative change of renormalized constants
with respect to the bare ones, eventually depends only on the \textit{initial} values $\varepsilon_w(0)/T$ and $L/ w(0)$. 
In other words, what matters are 
\begin{itemize} 
\item the temperature in units of the DW energy, $\varepsilon_w(0)$
\item and length scales in units of the DW width, $w(0)$. 
\end{itemize} 
Besides that, from Eq.~\eqref{Fogedby_eq} it is clear that $\Lambda/w$ only depends on $\varepsilon_w/T$. 
Therefore, applying the same argument as for the renormalization flux displayed in Fig.~\ref{Fig_12}, 
we expect  $\tilde{\Lambda}(T)/w(0)$  to be a function of $\varepsilon_w(0)/T$ only.
The correlation length is expected to fulfill the same scaling property as  $\tilde{\Lambda}$: 
$\xi/w(0)$ has to be a \textit{universal} function of $\varepsilon_w(0)/T$. 
In the following we will see that this property holds true provided 
that $\varepsilon_w(0)$ is replaced by the DW energy obtained from the discrete-lattice calculation $\varepsilon_{dw}$ 
(we remind that Fig.~\ref{fig1_intro} highlights some deviations of $\varepsilon_w$ from $\varepsilon_{dw}$  
in the vicinity of $D/J\!=\!2/3$). 
%%%%%%%%%%%%%%%%%%%%%%%%%%%%%%%%%%%%%%
\subsection{Transfer Matrix \textit{versus}\\Polyakov Renormalization}  
%
%%%%%%%%%%%%%%%%%%%%%%%%%%%%%%%
\begin{figure}
\begin{center}               
\includegraphics[width=7.5cm]{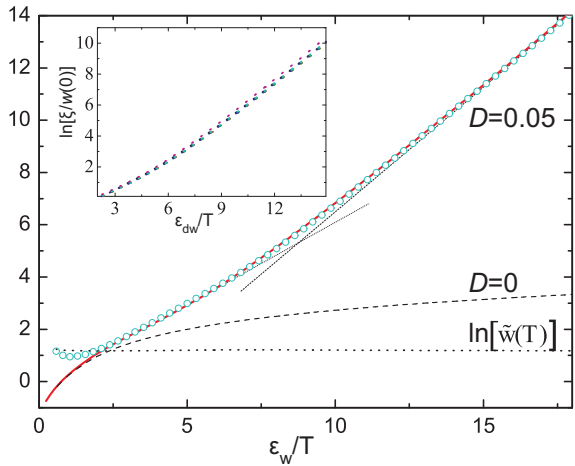}
\end{center}
\caption{Color online. Logarithm of the correlation length \textit{vs.} $\varepsilon_w(0)/T$: TM calculation for $D=0.05$ (solid red line); 
 $\tilde{\Lambda}(T)$ multiplied by a constant to match the TM results at low $T$ (circles); analytic result  
for $D=0$ given in Ref.~\onlinecite{Fisher63AJP} (dashed line). 
The straight lines are guide to the eye to highlight the change of slope due to spin-wave renormalization. 
The dotted line corresponds to the renormalized DW width $\ln[\tilde w(T)]$. 
Inset: Logarithm of $\xi/w(0)$  \textit{vs.} $\varepsilon_{dw}(0)/T$ for $D=$ 0.1, 0.2, 0.3, 0.4, 0.5, 0.6 (same data as in Fig.~\ref{fig1}).  
\label{Fig_13}}
\end{figure}
%%%%%%%%%%%%%%%%%%%%%%%%%%%%%%%%%
%
As stated before, the temperature dependence of the correlation length  
is expected to follow the behavior of $\tilde{\Lambda}(T)$.   
In Fig.~\ref{Fig_13}, we compare $\xi(T)$ computed by means of the TM technique with $\tilde{\Lambda}(T)$ as a function 
of $\varepsilon_w(0)/T$. We chose the specific value $D=0.05$ because for larger values of $D$ the correction 
to the DW energy due to the discreteness of the lattice is not negligible and the identification $\varepsilon_{dw}\!=\!2\sqrt{2DJ}$ is not totally justified 
(see Fig.~\ref{fig1_intro}). 
The values of $\tilde{\Lambda}(T)$ (crosses) have been shifted by a constant factor to match the TM results (line-symbols) at low  temperature. 
Indeed, the temperature behavior of $\tilde{\Lambda}(T)$ closely follows that of the correlation length till relatively high temperatures. 
The change in the slope occurring at intermediate temperatures is well reproduced by the RG calculation.  
This means that spin-wave renormalization is the main physical reason for the reduction of $\Delta_\xi$ with increasing temperature 
observed in the broad-wall regime (dotted lines in Fig~\ref{fig1}).   
In the RG language, such a change in the slope can also be interpreted as a crossover towards the temperature at which $\tilde{D}(T)$ vanishes.  
This phenomenon is the analogous of the reorientation transition in magnetic films~\cite{Pescia-Pokroksky}. 
The progressive vanishing of the uniaxial anisotropy with increasing temperature reflects in the fact that for the highest temperatures reported in Fig.~\ref{Fig_13} the TM calculation recovers the analytic solution for the isotropic Heisenberg chain, with $D(0)\!=\!0$ (solid line)~\cite{Fisher63AJP}.
In such a region, the behavior of $\tilde{\Lambda}(T)$ deviates from the one of the correlation length. However, 
at such short distances the RG treatment loses its meaning since the length scale at which the renormalization stops, $\tilde{\Lambda}(T)$, becomes of the order of the 
DW width $\tilde{w}(T)$ (dotted blue line) or even smaller.   \\
%%%%%
According to the RG analysis discussed in subsection~\ref{Symm_Ren_flux} one expects  
that the ratio $\xi/w(0)$ be given by a universal scaling function of $\varepsilon_w(0)/T$.  
However, the whole RG approach is based on the continuum approximation, Eq.~\eqref{Heisenberg_Ham_cont},   
which is expected to hold for $D\!\ll \!J$, $i.e.$ when DWs are significantly broad.   
But in fact, the discrepancies between $\varepsilon_w(0)\!=\!2\sqrt{2JD}$ 
and the corresponding values obtained from the discrete-lattice calculation,  $\varepsilon_{dw}$,   
are significant for most of the $D$ and $J$  used to produce the curves in Fig.~\ref{fig1}.   
For $D/J\! < \!2/3$ but not $D\!\ll\! J$, we can \textit{tentatively} extend the validity of the scaling property of $\xi$ 
beyond the continuum limit by replacing $\varepsilon_w$ with the DW energy obtained from the 
discrete-lattice calculation.  Therefore we propose that 
\begin{equation}
\label{scaling_ansatz}
\frac{\xi(T)}{w(0)} = f\left(\frac{\varepsilon_{dw}}{T}  \right)  
\end{equation}
with $f$ universal scaling function. 
In the inset of Fig.~\ref{Fig_13} we check the validity of Eq.~\eqref{scaling_ansatz} 
by plotting $\xi(T)/ w(0)$ as a function of the ratio  $\varepsilon_{dw}/T$ for $J\!=\!1$ and different values of 
$D\!=\! 0.1 - 0.6$.  
The correlation length data are the same as the ones plotted in Fig.~\ref{fig1}.  
The data collapsing predicted by Eq.~\eqref{scaling_ansatz} is indeed observed in the inset of Fig.~\ref{Fig_13}. 
Remarkably, the log-linear plot of the scaling function $f\left(\varepsilon_{dw}/T  \right)$ \textit{vs.}  $\varepsilon_{dw}/T$  does not show a constant slope  
as a consequence of the intrinsic temperature dependence of  $\Delta_\xi$ characterizing the broad-wall regime.   
%
%%%%%%%%%%%%%%%%%%%%%%%%%%%%%%%%%%%%%%%%%%%%%%%%%%%%%%%%%%%%%%%%%%
%
\section{Dynamic properties: TQMC\label{sec_IV}} 
In order to study the dynamics of the model defined by Eq.~\eqref{eq1}, we used a time-quantified Monte-Carlo algorithm (TQMC)\cite{Nowak00PRL,Cheng06PRL} 
fixing $N\!=\!100$ and implementing both periodic and open boundary conditions.  
In the TQMC scheme, Monte-Carlo steps (MCS) are mapped into real-time units through the following relation~\cite{Billoni07JMMM},
%
%%%%  EQUATION  %%%%%%%%%%%%%%%%%%%%%%%%%%%%%%%%%%%%%%%%%%%%%%
\begin{equation}\label{eq01}
\Delta t[\tau_K]= \frac{D}{40 T}R^2\,\Delta t[\mbox{MCS}]\,.  
\end{equation}
%%%%%%%%%%%%%%%%%%%%%%%%%%%%%%%%%%%%%%%%%%%%%%%%%%%%%%%%%%%%
\noindent This relation gives the variation of real time in units of the damping time $\tau_K$ 
where $R$ is the size of the cone used for updating single-spin configurations~\cite{Cheng06PRL,Garcia-Palacios98PRB}.  
Note that the factor that relates MCS with the real time is divided by the temperature $T$, therefore 
the time spanned in a given simulation increases when the temperature is decreased,
provided the factor $D R^2/(40T)$ is small enough.  
We set $R=0.025$ in all simulations. Thus, for  typical values
$D/J=0.1$ and $T/J=0.1$  one has  $D R^2/(40T)\sim 1.5\times10^{-5}$.  
On the other hand, the damping time is given by 
%
%%%%  EQUATION  %%%%%%%%%%%%%%%%%%%%%%%%%%%%%%%%%%%%%%%%%%%%%%%%%%%%%%%%%%%%%%%%%%%%%%%%%%%%%%%%%%%%%%%%%%
\begin{equation}\label{eq02}
\tau_K = \frac{1+a^2}{a}\frac{1}{\gamma \mu_0 H_K},  
\end{equation}
%%%%%%%%%%%%%%%%%%%%%%%%%%%%%%%%%%%%%%%%%
%
\noindent where $\mu_0 H_K$ is the anisotropy field, $a$ the adimensional damping constant and $\gamma$ the gyromagnetic
ratio~\cite{Billoni07JMMM}. 
Since the anisotropy field associated with Hamiltonian~\eqref{eq1}
%Since in a SCM the anisotropy field 
can be expressed as $\mu_0H_K = 2D/g\mu_B$ and $\gamma=g \mu_B/\hbar$,   
then $\gamma \mu_0 H_K =2D/\hbar$. Henceforth, the quantity $\hbar/2D$ will be assumed as time unit. 
In real SCMs $D$ is of the order $10$ K. 
We used an intermediate value for the damping constant, $a=0.25$, 
therefore $\tau_K$ is in the range of picoseconds ($\hbar=7.6$ K ps), with these parameters. 
Moreover, the maximum number of MCS that we spanned in our simulations is of the order of $10^8$ $\mbox{[MCS]}$, 
which would correspond to a total time $\sim 0.1 \mu$s for $D/J=T/J$. 
For TQMC simulations we set the following values for the ratio between the anisotropy and the exchange strength  
$D/J\! = \! 0.1, 0.2, 0.3, 1$ and 2. These ratios are comparable to values of real SCMs~\cite{Coulon04PRB,Miyasaka06CEJ} 
and correspond to  different regimes of the DW profile~\cite{Barbara94JMMM,Vindigni08ICA}  (see Fig.~\ref{fig1_intro}).  
We studied magnetic relaxation for the selected values $D/J=0.1, 0.2$ and $0.3$, all falling in the broad-wall regime. The  
temperatures were ranged between the values indicated in Table~\ref{tab1}.  
In relaxation simulations, such values of $T$ were chosen in order that the correlation length was smaller than the system size, even if for the 
lowest temperatures $\xi$ and $N\!=\!100$ are of the same order of magnitude.       
The applied field was chosen so that $H=0.05 D $ in every simulated relaxation.  
In the next section we will present results concerning a detailed study of the trajectory displayed  
by DWs because of thermal fluctuations. This investigation was also performed by means of TQMC simulations for different temperatures and anisotropy values. 
For the diffusion studies in the broad-wall regime we set $D/J\!=\!0.1$ and $D/J\!=\!0.2$; temperatures were chosen in the ranges  
$T/J\!=\!0.01-0.04$ and $T/J\!=\!0.014-0.056$, respectively (see Table~\ref{tab1}). 
For the studies in the sharp-wall limit we used $D/J\!=\!1$ and $2$ while temperatures were chosen in the ranges $T/J\!=\!0.1-0.25$
and $T/J\!=\!0.25-0.48$ respectively.  
Table \ref{tab1} summarizes the parameters and physical quantities related to the different types of simulations. 
\begin{table}%[h]
\caption{\label{tab1} Parameters and physical quantities. }
\begin{ruledtabular}
\begin{tabular}{lccccc}
\\[-3mm]
\vspace{1mm}
$D/J$  &  $\varepsilon_w/J$  &  $w$  &  $T/J$ range  & $\xi$ &  simulation  \\   
\hline \\[-3mm]
%____
0.1      &  0.894          &  2.23     &  0.116 - 0.26  & 86 - 5   & Relaxation  \\ 
0.2      &  1.264          &  1.58     &  0.153 - 0.3   & 89 - 5   & Relaxation  \\ 
0.3      &  1.549          &  1.29     &  0.181 - 0.33  & 83 - 5   & Relaxation  \\ 
0.1      &  0.894          &  2.23     &  0.01 - 0.04   & 86 - 5   & Diffusion   \\
0.2      &  1.264          &  1.58     &  0.014 - 0.056 & 89 - 5   & Diffusion   \\
1        &  $ 2$    	   &  1        &  0.1 - 0.25    & 7$\cdot10^7$ - 165   & Diffusion   \\
2        &  $ 4$	   &  1        &  0.25 - 0.48   & 536 - 10   & Diffusion   

\vspace{1mm}    \\
%__________________________________________________________________
\end{tabular}
\end{ruledtabular}
\vspace{1mm} \\
The DW width $w$ and the correlation length $\xi$ are given in lattice units. The latter has been rounded to integer values.
\end{table}
\subsection{Relaxation curves} 
\begin{figure}%[ht]
\includegraphics*[width=8cm,angle=0]{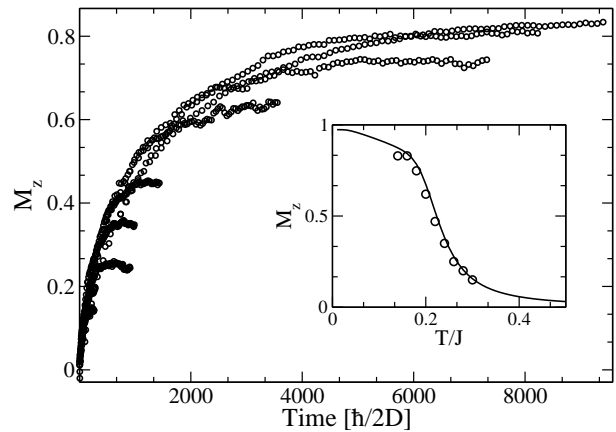}
\caption{\label{fig2}  Relaxation curves at different temperatures for $D/J=0.2$. Inset: 
FC curve obtained by TM calculations (solid line); open circles are the values of $M_0$ obtained by fitting the relaxation
curves with Eq.~\eqref{eq2} at different temperatures.
}
\end{figure}

\begin{figure}
\includegraphics*[width=8cm,angle=0]{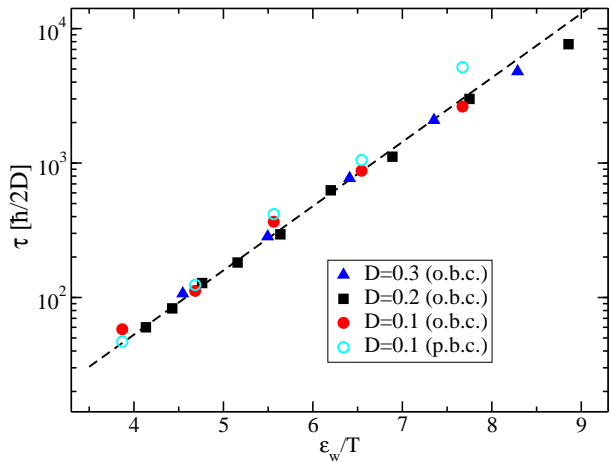}
\caption{\label{fig3} Color online. Arrhenius plot of the relaxation times obtained at different temperatures and
for three values of the anisotropy, $D/J=0.1$ (full circles), 0.2 (squares) and 0.3 (triangles), with open boundary conditions.  
The dashed line is guide to the eyes and has slope $1.1$.
Empty circles: periodic boundary conditions for $D/J=0.1$.}
\end{figure}

For studying how the magnetization relaxes towards equilibrium we used the following protocol. 
In each simulation, the system was cooled down in zero applied field at a 
cooling rate $r_0 = 5\cdot10^{-7}$; this means that the temperature was decreased according to the expression  $T = T_0 - r_0 t$, 
with $T_0/J = 1$ being the initial temperature  and  $t$ the time measured in MCS.  
Once the temperature at which we wanted to simulate relaxation was reached, we let the system equilibrate in zero field. 
Further on, we switched the field on and let the magnetization evolve towards its equilibrium value. 
The equilibration time in zero field was chosen to be equal 
to the relaxation time (in $H\!\ne\!0$).   
Fig.~\ref{fig2} shows the relaxation of the magnetization under a homogeneous applied field of strength $H/J=0.01$ along 
the $z$ direction for different temperatures. The anisotropy value is $D/J = 0.2$.  \\
All these curves can be fitted well with a stretched exponential, like in Ref.~\onlinecite{Coulon04PRB}: 
%
%%%%  EQUATION 2 %%%%%%%%%%%%%%%%%%%%%%%%%%%%%%%%%%%%%%%%%%%%%%%%%%%%%%%%%%%%%%
\begin{equation}\label{eq2}
M_z(t)=M_0\left[ 1-e^{-\left(\frac{t}{\tau}\right)^{\alpha}} \right] \,.
\end{equation}
%%%%%%%%%%%%%%%%%%%%%%%%%%%%%%%%%%%%%%%%%%%%%%%%%%%%%%%%%%%%%%%%%%%%%%%%%%%%%%%%%%%%%%%%%%%%%%%%%%%%%%%%%%%
%
\noindent For all the relaxation processes we found that $0.8\le \alpha \le 1$. 
%%% FIGURA 2 
%%%%%%%%%%%%%%%%%%%%%%%%%%%%%%%%%%%%%%%%%%%%%%%%%%%%%%%%%%%%%%%
%%%%%%%%%%%%%%%%%%%%%%%%%%%%%%%%%%%%%%%%%%%%%%%%%%%%%%%%%%%%%%%%%%%%%%%%%%
\noindent 
In the inset of Fig.~\ref{fig2} we report field-cooling curve (FC) as a function of $T/J$ obtained by TM calculations 
for an infinite system (full continuous line). The applied field is the same as in relaxation simulations, $H/J=0.01$. 
The points represent the values of $M_0$ obtained from fitting the computed relaxation curves 
with Eq.~\eqref{eq2}.  The very good agreement confirms that the 
magnetization indeed relaxed to its equilibrium value in all the simulated relaxation processes. 
\noindent 
Fig.~\ref{fig3} shows the relaxation times, $\tau$, obtained by fitting different relaxation curves, analogous to the ones shown in Fig.~\ref{fig2},  
for three different values of the anisotropy $D/J\!=\!$ 0.1, 0.2 and 0.3. The relaxation times are plotted in 
a log-linear scale as function of the inverse temperature normalized to the corresponding DW energy. 
An Arrhenius dependence of $\tau$ on the temperature is indeed evidenced for any value of $D$ and for both open and 
periodic boundary conditions (o.b.c. and p.b.c. respectively).  
The values of $\tau$ obtained with o.b.c lie on the same line whose slope is 
roughly 1.1. The same calculation was repeated with p.b.c. for $D/J\!=\!0.1$ only. 
The corresponding points (empty  circles in Fig.~\ref{fig3})  can be assumed to lie on the same line as for o.b.c. up to the value $\varepsilon_w/T\sim 6.5$. 
The last point at lower temperature deviates significantly from the dashed line with slope 1.1.  
At this temperature, the correlation length is comparable to the system size, which explains the discrepancy between the calculation 
with p.b.c and o.b.c. (see Table~\ref{tab1}). More specifically, it is $\xi\! = \!N\! = \!100$ for  $\varepsilon_w/T \!=\! 7.86$.   
On the other hand, for $\varepsilon_w/T \!< \!6.5$ the relaxation time computed with open and periodic boundary conditions 
depends on the temperature likewise. This fact suggests that for such temperatures the relaxation is not affected by the 
details of boundary conditions, namely it is a bulk process.  \\
\noindent 
We cannot provide any trivial explanation for the {\it universal} slope observed for $\varepsilon_w/T \!< \!6.5$: $\Delta_{\tau} = 1.1 \varepsilon_w$. 
We remark that in the sharp-wall limit  the formula given in Eq.~\eqref{Delta_tau},  $\Delta_\tau = 2\Delta_\xi+ \Delta_A$, 
has been confirmed by a number of experiments on SCMs~\cite{Coulon04PRB,Miyasaka_review,Coulon06Springer,Bogani_JMC_08}. 
Due to the fact that for sharp DWs $\Delta_\xi\!=\!\varepsilon_w$ (with $\varepsilon_w \!=\!2J$), 
the same formula adapted to the broad-wall limit would predict a slope of about $2 + D/\varepsilon_w $.    
The value obtained by fitting the data in Fig.~\ref{fig3} 
is therefore about 50\% smaller than what would predict  
Eq.~\eqref{Delta_tau}. We will come back to this important point when comparing our results with the 
few available experimental data on SCMs in the broad-wall limit. 
%
%%%% FIGURA 3 %%%%%%%%%%%%%%%%%%%%%%%%%%%%%%%%%%%%%%%%%%%%%%%%%
%%%%%%%%%%%%%%%%%%%%%%%%%%%%%%%%%%%%%%%%%%%%%%%%%%%%%%%%%%%%%%%%
%
\section{Domain-wall diffusion\label{sec_V}}
In this section we discuss the kind of trajectory displayed by DWs at finite temperature.
Particular attention will be given to the temperature dependence of the diffusion coefficient both in the broad- and 
sharp-wall limit. For the latter case we will provide a numerical confirmation of the 
phenomenological law $D_s\sim e^{-D/T} $ (see Eq.~\eqref{activation_xi-tau-D}) which 
-- to the best of our knowledge --  was still missing in SCM literature. 
The Arrhenius-like dependence of the diffusion coefficient  
highlights that each elementary move of a DW occurs -- in the average -- through a   
thermally activated mechanism, in the sharp-wall limit. We will show that this is not the case in the broad-wall regime.  
%  
%
%%%%%%%%%%%%%%%%%%%%%%%%%%%%%%%%%%
\subsection{Analysis of domain-wall trajectories}  
As already mentioned, the assumption that DWs perform a random walk induced by thermal fluctuations is 
the basic ingredient to relate the correlation length to relaxation time. 
Such assumption can be verified directly in TQMC by analyzing the microscopic configurations 
explored during a simulation. Details about this analysis can be found in Appendix~\ref{App_DW_traj}.
The outcome is the average trajectory followed by each DW at a given temperature.  
We found that, the diffusion relation $\langle x^2\rangle \propto t$ is not obeyed at short times (see Eq.~\eqref{eq3}).  
However, DW trajectories could be fitted well with a more general expression which describes a random walk with correlated steps~\cite{Taylor20PLMS}: 
%
%%%%  EQUATION 4 %%%%%%%%%%%%%%%%%%%%%%%%%%%%%%%%%%%%%%%%%%%%%%%%%%%%%%%%%%%%%%%%%%%%%%%%%%%%%%%%%%%%%%%%%%
\begin{equation}\label{eq4}  
\langle x^2 \rangle  = \sigma^2 \left( \frac{t}{\tau_c} + \frac{e^{-\frac{2t}{\tau_c}} -1}{2} \right) \, . 
\end{equation}
%%%%%%%%%%%%%%%%%%%%%%%%%%%%%%%%%%%%%%%%%%%%%%%%%%%%%%%%%%%%%%%%%%%%%%
%
\noindent $\tau_c$ is the characteristic crossover time from the ballistic regime at
\textit{short} times to the diffusive regime at \textit{longer}  times. When $t \ll \tau_c$ we are in the regime of 
correlated steps. Expanding Eq.~\eqref{eq4} accordingly, for $t/\tau_c \ll 1$, one obtains the  
ballistic relation between the displacement and time: 
\begin{equation}  
\langle x^2 \rangle  = \frac{\sigma^2}{\tau_c^2}t^2\,. 
\end{equation}
\noindent When $t \gg \tau_c$, the diffusion equation is recovered:  
\begin{equation}  
\label{long_times_diff}
\langle x^2 \rangle  = \frac{\sigma^2}{\tau_c} t =2D_st, 
\end{equation}
\noindent with $D_s$ being the diffusion coefficient ($2D_s\!=\!\sigma^2/\tau_c$). 
By fitting the mean-square displacement of DWs with Eq.~\eqref{eq4} both the diffusion coefficient 
and the crossover time, $\tau_c$,  can be obtained. 
In Table~\ref{tab3} we report the values of such parameters for different temperatures, 
$D/J\!=\!0.2$ and $H\!=\!0$.  
Note that both $D_s$  and $\tau_c$ decrease when the damping constant increases.
For low applied fields, like those used in relaxation simulations, $D_s$ and  $\tau_c$ are 
independent of the field itself.  
\begin{table}
\caption{\label{tab3}Diffusion  parameters.}
\begin{ruledtabular}

\begin{tabular}{lllcll}
%_________________________________
\\[-3mm]
\vspace{1mm}
$T/J$   &  $a$  &  $\tau_c$ & $\sigma$ & $2D_s$  \\   
\hline \\[-3mm]
%_________________________________
0.14       &  0.25   &  85.0   & 4.04   & 0.44  \\
0.18       &  0.25   &  16.7   & 2.01   & 0.49  \\
0.22       &  0.25   &  3.97   & 1.10   & 0.52  \\ 
0.14       &  4.00    &  32.3  & 2.01   & 0.35  \\ 
0.18       &  4.00    &  7.00  & 1.09   & 0.41  \\
0.22       &  4.00    &  2.12  & 0.66   & 0.45   

\vspace{1mm}    \\
%_________________________________
\end{tabular}
\end{ruledtabular}
\vspace{1mm} \\
\end{table}

In conclusion, over some time window -- generally larger the lower the temperature is --
each DW performs a ballistic motion before the genuine diffusion process starts. 
\subsection{Temperature dependence of the diffusion coefficient}
\begin{figure}
\includegraphics*[width=8cm,angle=0]{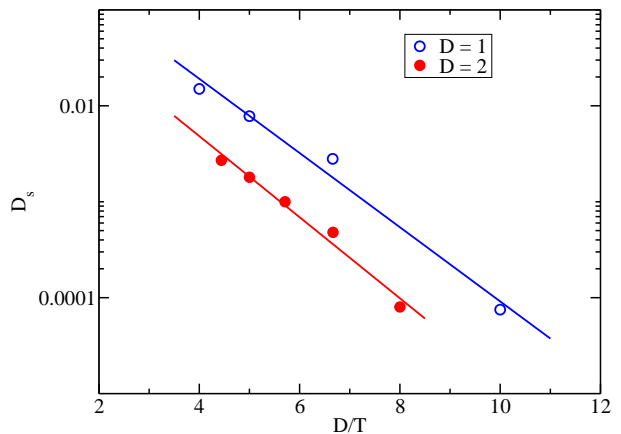}
\caption{\label{fig9} Color online. Temperature dependence of the diffusion coefficient for $D\!=\!$1 and 2 (sharp-wall regime). }
\end{figure}

\begin{figure}
\includegraphics*[width=8cm,angle=0]{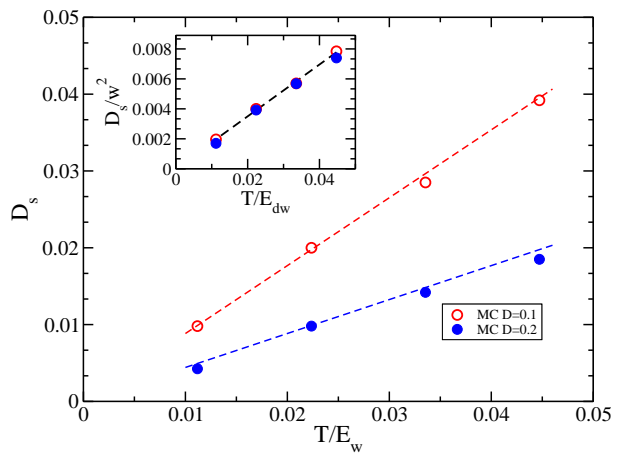}
\caption{\label{fig10} Color online. Temperature dependence of the diffusion coefficient for two values of the anisotropy 
$D=0.1$ and 0.2 (broad-wall regime). The inset shows the scaled curves. }
\end{figure}

In order to obtain quantitative results about the temperature dependence of diffusion coefficient $D_s$ 
(see Eq.~\eqref{long_times_diff}) we introduced a DW at the center of the spin chain with anti-periodic boundary conditions
(the spins at each end were kept anti-parallel to each other along the $z$ axis: $S_1^z=1$ and $S_N^z=-1$). 
In each numerical experiment, we thermalized the system at a given temperature and then followed the 
DW trajectories.  
The regimes with $D/J\!>\!2/3$ and $D/J\!<\!2/3$  will be analyzed separately. \\
%%%%%%%%%%%%%%%%%%%%%%%%%%%%%%%%%%
\textit{Sharp-wall regime} -- 
In Fig.~\ref{fig9} we plot in a log-linear scale the diffusion coefficient as a function of  $D/T$ 
for anisotropy $D/J\!=\!1$ and 2.  
In this scale the points simulated can be fitted well by a straight line indicating an Arrhenius behavior, $D_s\!\sim \! e^{-\Delta_A/T}$.  
This fact confirms the validity of the expression given in Eq.~\eqref{activation_xi-tau-D} for the diffusion coefficient in the sharp-wall regime:  
the slope is $\Delta_A \!=\! 0.89 D$ and  $\Delta_A \!=\! 0.97 D$ for   $D/J\!=\!1$ and 2, respectively.
The slight reduction of the energy barrier with respect to the prediction of Eq.~\eqref{activation_xi-tau-D} 
($\Delta_A\!=\!D$) may be due to the high-temperature points in Fig.~\ref{fig9}. 
Even for the attempt frequency of a single nanoparticle with uniaxial anisotropy~\cite{Brown63PR}, an Arrhenius behavior 
is expected \textit{only} in the limit $D/T \! \ll \!1$. 
Another possibility is that $\Delta_A$ becomes smaller than $D$ as 
the crossover ratio, $D/J\!=\!2/3$, is approached.  
In this sense one would expect the renormalizing effect of spin waves to be more important for $D/J\!=\!1$  than for  $D/J\!=\!2$  (see  Sec.~\ref{sec_III}). 
Therefore, within the numerical accuracy, our TQMC simulations confirm the phenomenological law proposed in  Ref.~\onlinecite{Coulon04PRB,Kirschner97} with $\Delta_A\!=\!D$. 
%
%
%%%% FIGURA 9 %%%%%%%%%%%%%%%%%%%%%%%%%%%%%%%%%%
%%%%%%%%%%%%%%%%%%%%%%%%%%%%%%%%%%%%%%%%%%%%%%%
%%%%%%%%%%%%%%%%%%%%%%%%%%%%%%%%%%
\textit{Broad-wall regime} -- 
The temperature dependence of the diffusion coefficient changes when the anisotropy-to-exchange ratio is reduced. 
In Fig.~\ref{fig10} we plot the diffusion coefficient {\it vs. } temperature for the two ratios $D/J\! =\! 0.1$ and $0.2$ (broad-wall regime).  
The temperature is now expressed in units of the DW energy $\varepsilon_{w}$, given in Eq.~\eqref{definition_E_w}.  
In this case the diffusion coefficients increase {\it linearly} with temperature.
This behavior is at odds with the Arrhenius dependence predicted by Eq.~\eqref{activation_xi-tau-D} for sharp DWs: 
% but  it is more reminiscent of the one observed for a massive particle diffusing in viscous medium. 
it rather reminds the behavior of the diffusion coefficient of a massive particle in a viscous medium. 
The slope of the diffusion coefficient as a function of  $T/\varepsilon_w$ is larger the smaller the anisotropy is.  
%    
%%%% FIGURA 9 %%%%%%%%%%%%%%%%%%%%%%%%%%%%%%%%%%%%%%%%%%%%%%%%%%%%%%%%%%%%%%%%%
%%%%%%%%%%%%%%%%%%%%%%%%%%%%%%%%%%%%%%%%%%%%%%%%%%%%%%
Following analogous considerations to those that allowed us to derive the scaling relation in Eq.~\eqref{scaling_ansatz}, 
we can \textit{attempt} to propose a scaling ansatz for the diffusion coefficient. Note that the units of $D_s$ are 
square unit lengths divided by a unit time. The time unit assumed throughout the paper is  $(\gamma_0 \mu_0 H_k)^{-1}\!=\! \hbar/2D$.
This unit corresponds, in general, to different physical time scales 
for different values of $D$. Nevertheless, this specific 
dependence on $D$ has been already eliminated from $D_s$ by measuring the time in units  $\hbar/2D$.
Then we need to express the lengths in unit of $w$ and the energies in units $\varepsilon_{dw}$, which yields 
\begin{equation}
\label{scaling_ansatz_D_s}
\frac{D_s(T)}{w^2} = g\left(\frac{\varepsilon_{dw}}{T}  \right)  \,.
\end{equation} 
From Fig.~\ref{fig10} and from the analogy with a particle in viscous medium, we conclude that 
the scaling function $g$ is just a line passing through the origin in the $T$-$D_s$ plane so that 
\begin{equation}
\label{scaling_ansatz_D_s_1}
\frac{D_s(T)}{w^2} = A \frac{T}{\varepsilon_{dw}} \,,
\end{equation} 
$A$ being some constant with units  $2D/ \hbar$. 
In the inset of Fig.~\ref{fig10} we plot $D_s/w^2$ 
\textit{vs.} $T/\varepsilon_{dw}$ using the same data as in the main frame. 
Indeed, the scaling prediction of Eq.~\eqref{scaling_ansatz_D_s_1} is well-obeyed, giving $A=0.17$ $[2D/ \hbar]$. 
\section{Phenomenological arguments\label{sec_VI}} 
The analysis performed for both static and dynamic properties allows stating that
the relevant energy scales in our problem 
are the DW energy $\varepsilon_w$ and the {\it width} and {\it position} of the spin-wave spectrum.    
Such energies, and their relationship with the thermal energy, determine the physics of SCMs described 
by the model Hamiltonian~\eqref{eq1}.  
The full spin-wave spectrum can be obtained by linearizing the
Landau-Lifshitz equation corresponding to Hamiltonian~\eqref{eq1}. 
The energy of a spin wave with frequency $\omega$ and wave vector $q$ is 
\begin{equation}
\label{dispersion_relation_sw} 
\hbar \omega(q)  = \left| S^z  \right| 
\left[2J \left(1- \cos(q)\right) +2D  \right]\,.
\end{equation}
At low enough temperature, in a spatial region delimited by two DWs one essentially has $\left| S^z  \right|=1$. 
%
%%%%%%%%%%%%%%%%%%%%%%%%%%%%%%%%%%%%%%%%%%%%%%%%%%%%%%%%%%%%%%%%
\begin{figure}[!b]
\includegraphics[width=8.6cm]{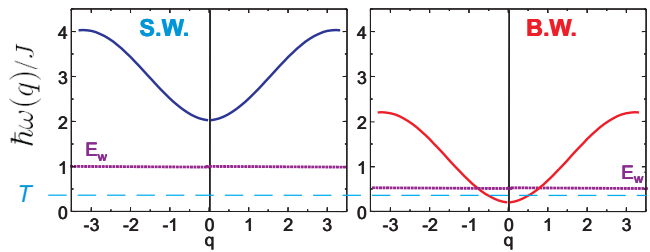}
\caption{\label{DW_SW_energy} Color online. The solid lines represent the energy of spin waves $\hbar\omega(q)$ (in $J$ units) 
as a function of the wave vector $q$. Horizontal lines
correspond to the energy of one DW in the sharp- and in the broad-wall regime: 
for the left panel $D/J\!=\!2$ while for right panel $D/J\!=\!0.2$. 
The horizontal dashed line represents an indicative temperature smaller than $\varepsilon_w$ in both cases (see the text). }
\end{figure}
%%%%%%%%%%%%%%%%%%%%%%%%%%%%%%%%%%%%%%%%%%%%%%%%%%%%%%%%%%%%%%%%
%
In the left panel of  Fig.~\ref{DW_SW_energy} the spectrum of fluctuations, $\hbar\omega(q)$,
is plotted \textit{vs.} $q$ for $D \!=\! 2.5 J$  (the sharp-wall limit).
The dashed horizontal line indicates the corresponding DW energy $\varepsilon_w\!=\!2 J$. 
Clearly the energies of the two family of excitations -- spin waves and DWs -- 
are well separated from each other. 
Moreover, the genuine 1d character of a spin chain is evident
when the energy of thermal fluctuations is lower than the DW energy:  $ T < \varepsilon_w$.
At these temperatures some short-range correlations develop, $i.e.$, $\xi$ exceeds some lattice units. 
The dotted horizontal line highlights a realistic reference for such a thermal energy. 
On the right panel, the same plot is displayed for $D\!=\!0.2 J$  corresponding
to broad DWs. In this case the dashed horizontal line, representing the DW energy $\varepsilon_w$, passes through the
spectrum of spin waves. As a consequence, in the broad-wall regime
one expects the interplay between DWs and spin waves to affect crucially the finite-temperature properties of the system.  
On the other hand, as in Fig.~\ref{DW_SW_energy} (left) the spin-wave spectrum lies well above $\varepsilon_w$ and $T$,
spin waves are expected to play no essential role in sharp-wall limit (when $T\!<\!\varepsilon_w\!<\! 2D$). 
The fact that $\Delta_{\xi}$ computed in Sect.~\ref{sec_III} is independent of $T$ for sharp DWs while it effectively depends on temperature for broad DWs, 
confirms the heuristic argument evidenced by Fig.~\ref{DW_SW_energy}. In particular,  we have shown 
through Polyakov renormalization that the interplay between spin waves and DWs at finite temperature gives 
a quantitative explanation for the dependence of $\Delta_{\xi}$ on $T$ in the broad-wall limit. 
Parenthetically, we note that in the time domain it is easy to see that averaging over spin waves 
(the scalar fields $\phi_a$ in the language of Polyakov renormalization) corresponds to 
an integration over  \textit{fast} fluctuations. In fact, 
according to Eq.~\eqref{dispersion_relation_sw} the typical time periods of spin waves, $2\pi/\omega(q)$, 
are of the order of our reference time unit $\hbar/2D$  or even smaller.   
The typical time scale for the creation or annihilation of DWs is, instead, of the order of the relaxation time $\tau$, $i.e.$, several orders of magnitude 
larger than $\hbar/2D$.   Thus, in the experimental situation relevant for SCMs it is clearly $\tau\!\gg\!1/\omega(q)$, meaning that   
$\vec{n}(x)$ represents a {\it slow} varying field and $\vec{\phi}(x)$ a {\it fast} varying field. \\
For what concerns the relaxation time we cannot provide an effective argument as Polyakov renormalization to justify 
our numerical findings. However, it seems reasonable that the very same interplay between DWs and spin waves 
affects the temperature dependence of the relaxation time in a similar way as it affects the correlation length. 
A natural consequence of this is that Eq.~\eqref{Delta_tau} does not necessarily hold true in the broad-wall limit.    
One has to be very cautious even in trying to generalize the relation 
$\Delta_\tau = 2\Delta_\xi+ \Delta_A$ (Eq.~\eqref{Delta_tau}) to the case of broad DWs. 
In this regard, we recall that for $D/J\!<\!2/3 $
\begin{enumerate}  
\item $\Delta_{\xi}$ depends on the temperature range in which it is measured
\item the time window in which the DW motion is ballistic, and not diffusive, becomes larger while lowering the temperature
\item in the temperature range that we investigated with TQMC it is $\Delta_A\!=\!0$, meaning that each single DW move is not thermally activated. 
\end{enumerate}  
For $4 \,\varepsilon_w \!<\!T\!<\! 9\, \varepsilon_w$ and $\xi\!<\!N$, the numerical  data reported in Fig.~\ref{fig3} 
suggest that Eq.~\eqref{Delta_tau} has to be modified into 
\begin{equation} 
\label{empirical_Delta_tau} 
\Delta_\tau = 1.1 \, \varepsilon_w
\end{equation} 
for broad DWs. In order to check this prediction against experiment 
we refer to  two Mn(III)-based spin chains with $D/J\!<\!2/3$ (broad DW) reported in Ref.~\onlinecite{Miyasaka06CEJ,Balanda}.  
Both chains are better described by the Seiden model~\cite{Seiden}  with anisotropy (on the classical spin) rather than the Heisenberg model in Eq.~\eqref{eq1}. In fact,
the Mn centers ($S\!=\!2$) alternate with an organic radical (TCNE or TCNQ) whose magnetic contribution is essentially the same 
as a free electron: spin $s\!=\!1/2$ and Land\'e factor $g_s\!=\!2$.    
In the broad-wall limit, the Seiden model with anisotropy can be mapped into the Heisenberg model described by Hamiltonian~\eqref{eq1} with an halved 
exchange coupling~\cite{Coulon_private}. 
For the Mn(III)-TCNE spin chain~\cite{Balanda}, we have estimated in Sect.~\ref{sec_III} $\varepsilon_w\simeq100$ K.
Adapting to our convention the values of $D$ and $J$  given in Ref.~\onlinecite{Miyasaka06CEJ} we obtain 
$\varepsilon_w\simeq 86$ K for the Mn(III)-TCNQ spin chain. 
Thus Eq.~\eqref{empirical_Delta_tau} would predict the following activation barriers for relaxation 
\begin{itemize}  
\item for the Mn(III)-TCNE spin chain $\Delta_\tau \! \simeq \! $ 110 K, to compare with the experimental value  $\Delta_\tau^{exp} \!= \! $ 117 K 
\item for the Mn(III)-TCNQ spin chain $\Delta_\tau \! \simeq \! $ 95 K, to compare with the experimental value  $\Delta_\tau^{exp} \!= \! $ 94 K.
\end{itemize}  
The prediction of Eq.~\eqref{empirical_Delta_tau}
agrees well with the measured energies in both cases. 
We note, however, that in experiments the validity of the empirical formula $\Delta_\tau \!= \!1.1\, \varepsilon_w$ seems to extend 
down to a temperature region in which $\xi\! > \!N$. 
In fact, the lowest  temperature for which the  $\Delta_\tau^{exp} \!=\!94$ K for the Mn(III)-TCNQ spin chain~\cite{Miyasaka06CEJ} 
is roughly $T\!=\!4.5$ K. The corresponding correlation length, extrapolated from Fig.~\ref{fig1}, should be of the 
order of $10^7$ Mn(III)-TCNQ units. The same estimate gives a correlation length of the order of $10^{12}$ units for 
the Mn(III)-TCNE spin chain~\cite{Balanda}. As already stated in Sect.~\ref{sec_III}, defects and dislocations typically limit 
the length of spin chains to $10^2-10^4$ units (see the dotted horizontal lines 
in Fig.~\ref{fig1}). Therefore, at the lowest temperatures at which Eq.~\eqref{empirical_Delta_tau} seems to apply, 
$\xi$ should exceed the average distance between two defects in both molecular spin chains.
A conclusive analysis would require a more accurate fitting of the model parameters to the experimental data for each sample.  
At this stage, we have no qualitative explanation nor a numerical confirmation for the validity of   
Eq.~\eqref{empirical_Delta_tau} in the regime $\xi\! > \!N$. 
Simulating a relaxation experiment at lower temperatures,  in the region where $\xi\! \gg \!N$, 
is computationally very expensive due to the Arrhenius dependence of the relaxation time.   
This issue, indeed, deserves further theoretical investigation but this is beyond the scope of the present work. 
\section{Conclusions}
We studied a model paradigmatic for classical spin chain or magnetic nanowires with uniaxial anisotropy and 
identified two distinct regimes for static and dynamic properties.
Such differences in the finite-temperature behavior are closely related to the thickness of DWs at 
zero temperature. We distinguished, accordingly, between the sharp- and broad-DW regimes. 
In the sharp-wall regime ($D/J\! > \!2/3$) the correlation length obtained by TM calculations 
shows an activated behavior as function of the inverse of the temperature.
The corresponding activation energy is equal to the DW energy $\Delta_\xi\! =\! \varepsilon_{w}$  ($\varepsilon_{w}\!=\!2J$ in this regime).
In fact, for large anisotropy-to-exchange ratios, DWs extend only over one lattice spacing so that 
the anisotropy energy does not affect two-spin correlations.  
At variance, when DWs develop over several lattice units ($D/J\!<\!2/3$), 
the correlation length still shows an exponential divergence with the inverse of the temperature, but with  
a temperature-dependent $\Delta_\xi $. At low temperatures the activation energy is larger   
$\Delta_\xi \!= \!0.9 \varepsilon_{w}$ (here $\varepsilon_{w}\! =\! 2\sqrt{2DJ}$), whereas at higher temperatures 
it is $\Delta_\xi = 0.6 \varepsilon_{w}$. The first result agrees with a
low-temperature expansion available in the literature~\cite{Fogedby84JPCSSP}. 
Besides that, we provided a physical argument -- based on Polyakov renormalization -- to justify the 30\% of reduction of $\Delta_\xi $ observed at high temperature. 
This allowed us to conclude that the  
appearance of the lower activation energy at higher temperatures is due to the interplay between DWs and spin waves. 
This interplay is not significant in the sharp-wall regime where static properties are practically determined by the energy cost to create a DW at zero temperature.  
The reason why physics is remarkably different in the sharp- and broad-wall regime 
lies on the relative difference among the energy scales involved in the problem and the thermal energy     
corresponding to temperatures at which short-range correlations develop (see Sect.~\ref{sec_VI}).  

In SCMs, relaxation is usually assumed to be driven by DWs diffusion~\cite{Coulon04PRB,Miyasaka_review,Coulon06Springer,Bogani_JMC_08}.  
Based on this assumption,  
the activation energy for the correlation length,  $\Delta_\xi$,  and that of the relaxation time, $\Delta_\tau$, 
are then related with each other. 
We tested, with TQMC simulations, that DWs indeed perform a random walk for time intervals much longer than the typical precessional time of a single spin 
and determined the temperature dependence of the diffusion coefficient. 
In the sharp-wall regime, we found that the diffusion coefficient $D_s$ follows 
an Arrhenius behavior with an activation energy close to the anisotropy value, $D$, as assumed in most of the 
experimental works~\cite{Coulon04PRB,Kirschner97,Miyasaka_review,Coulon06Springer,Bogani_JMC_08}. 
In the broad-wall limit, the diffusion coefficient does not follow an activated mechanism but it rather grows linearly with the temperature, reminding  
the behavior of a particle in a viscous medium. 
The results of this analysis confirm the robustness of the random-walk argument relating the correlation length to the relaxation time~\cite{Tobochnik,Luscombe_PRE_96,daSilva_PRE_95,Barma_J_Stat_Phys_80} 
and suggest a dynamic critical exponent $z\!=\!2$.  
Nevertheless, in the broad-wall regime, the relation between the $\Delta_\xi$ and $\Delta_\tau$ is not trivial 
due to spin-wave renormalization. As a consequence, the joint theoretical and experimental characterization of SCMs falling in this regime cannot be based on the simple   
Glauber model~\cite{Glauber63JMathP} or generalizations of 
it~\cite{PRL_Bogani,Coulon04PRB,Vindigni_JMMM_04,Kirschner97,Coulon_PRB_07,Pini-Rettori_PRB_07,Pini-Vindigni_JPCM_09,Caneschi02EPL,Bernot_PRB_09}.

The symmetry of renormalization-group equations suggests  
the existence of scaling laws specific to the broad-wall regime:  the natural unit for 
length scales is the DW width $w\! =\! \sqrt{J/2D}$  while for energies it is $\varepsilon_{w}\! =\! 2\sqrt{2DJ}$ (the DW energy at zero temperature).  
Thus, one expects that the physical observables which depend only on these quantities be universal functions of 
properly rescaled variables (e.g. $\varepsilon_{w}/T$). 
Static TM calculations and dynamic TQMC simulations confirm the validity of this conjecture 
for the correlation length, the diffusion coefficient of DW motion and the relaxation time (for the latter, 
scaling is obeyed apart from a residual temperature dependence in the time unit intrinsic of the TQMC method).

In view of possible magneto-storage applications, increasing the thermal stability of the SCMs would be 
desirable. To this aim, we note that designing novel compounds with a larger $J$ as possible would not be a good strategy for two reasons: 
for $D/J \!< \!2/3$ i) the DW energy is the sole quantity which controls thermal stability  and it scales as $\varepsilon_{w} \! \sim \! \sqrt{J}$, instead of $\sim\!J$ like for sharp DWs; ii) spin-waves renormalization progressively lowers    
the effective energy barrier for relaxation as the temperature is increased 
(by renormalizing $\varepsilon_{w}$).     

Temperature is often neglected in models employed to study the current-induced DW motion in magnetic nanowires~\cite{Marrows_AdvPhys_05,Thiaville_EPL_05,view_point_Klaeui} or it is 
taken into account in  the phenomenological parameters of the Landau-Lifshitz-Gilbert equation~\cite{Nowak_PRB_09}. 
The basic hypothesis is that the considered nanowire behaves as a 3d magnet below its critical temperature~\cite{Ar_Abanov_PRL_10a,Ar_Abanov_PRL_10b}. 
However, recent experiments~\cite{Yamaguchi_APL_05,Junginger_APL_07,Franken_APL_09} suggest 
that Joule heating may induce the formation of domains in the nanowires, which  
highlights the restorations of a genuine 1d magnetic character. 
In this situation,  
the DW trajectory may result from a delicate combination of  
thermal diffusion (stochastic) and the deterministic motion induced by the electric current.     
Our study of the temperature dependence of the diffusion coefficient can be considered a preliminary
contribution to this problem, of technological relevance~\cite{Allenspach_PRL_09,Parkin11042008}, that indeed deserves further investigation.   
Typically in metallic nanowires (Co, Ni, Fe, Permalloy)  
$D\!\simeq\!1-10$ K ($\sim\!0.1-1$ meV) and  $J\!\simeq\!100-500$ K ($\sim\!10-50$ meV), meaning that they generally fall in the broad-wall regime 
where the temperature dependence of any observable is expected to be affected by the non-trivial interplay between DWs and spin waves. 
\begin{acknowledgments}
A. V. would like to thank Claude Coulon and Rodolphe Cl\'erac for drawing this problem under 
his attention and for fruitful discussions. Hitoshi Miyasaka is also acknowledged for 
sharing with us unpublished experimental results on broad-wall SCMs. 
We acknowledge the financial support of ETH Zurich and the Swiss National Science Foundation.
\end{acknowledgments}

\appendix
\section{The transfer-matrix approach\label{TM_Appendix}}
Given a general classical spin-chain  Hamiltonian with nearest-neighbor interactions 
\begin{eqnarray}
\label{E:NN_hamiltonian}
\mathcal{H}
=-\sum^{N}_{i=1}V(\vec{S}_{i},\, \vec{S}_{i+1})
\end{eqnarray}
the partition function $\mathcal{Z}$ is given by 
\begin{eqnarray}
\label{zeta_TM_general}
\mathcal{Z}=&\int d\Omega_{1}\int			\nonumber
d\Omega_{2}\ldots \int e^{\beta V(\vec{S}_{1},\, \vec{S}_{2})}\,
e^{\beta V(\vec{S}_{2} \vec{S}_{3})} \\ & \, \ldots \, e^{\beta
V(\vec{S}_{N},\, \vec{S}_{1})}\, d\Omega_{N}
\end{eqnarray}
where the integrals, $d\Omega_i$, are performed over any possible direction of the unit vectors $\vec{S}_{i}$ and $\beta=1/T$.
Defining the transfer kernel $\mathcal{K}$~\cite{Wyld} as 
\begin{eqnarray}
\label{TM_kernel}
\mathcal{K}(\vec{S}_{i},\, \vec{S}_{i+1})=e^{\beta
V(\vec{S}_{i},\, \vec{S}_{i+1})}
\end{eqnarray}
and taking periodic boundary conditions $(\!N\!+\!1\!=\!1\!)$, the partition
function $\mathcal{Z}$ takes the form of the trace of $N$-th power
of $\mathcal{K}(\vec{S}_{i},\, \vec{S}_{i+1})$:
\begin{eqnarray}
\label{TM_kernel_trace}
\mathcal{Z}&=\int d\Omega_{1}\int			\nonumber
d\Omega_{2}\ldots \int \mathcal{K} (\vec{S}_{1},\, \vec{S}_{2})\,
\mathcal{K} (\vec{S}_{2},\, \vec{S}_{3})\\ & \ldots \,
\mathcal{K}(\vec{S}_{N},\, \vec{S}_{1})\, d\Omega_{N}= \rm Tr\big\{
\mathcal{K}^{N}\big\} \,.
\end{eqnarray}
The computation of such a trace, as well as other physical observables, 
is simplified if one first solves the following  integral
eigenvalue problem:
\begin{eqnarray}
\label{TM_integral_eq}
\int
\mathcal{K}(\vec{S}_{i},\,\vec{S}_{i+1})\psi_{n}(\vec{S}_{i+1})
d\Omega_{i+1} =\lambda_{m} \psi_{m}(\vec{S}_{i})
\end{eqnarray}
The eigenfunctions fulfill the properties
\begin{eqnarray}
\label{completeness}
\sum_{m}\psi_{m}(\vec{S}_{i})\psi _{m}(\vec{S}_{j})=\delta
(\vec{S}_{i}-\vec{S}_{j}) \, \, \, \, \, \, 
\text{completeness}
\end{eqnarray}
and 
\begin{eqnarray}
\label{ortho} \int\psi_{n}(\vec{S})\psi _{m}(\vec{S})\,
d\Omega=\delta _{n,m}\, \, \, \, \, \, \, \, \, \,\, \, \, \, \,
\text{orthonormality}
\end{eqnarray}
where $\delta (\vec{S}_{i}-\vec{S}_{j})$ is the Dirac
$\delta$-function and $\delta_{n,m}$ is the Kronecker symbol. 
For non-symmetric kernels, $\mathcal{K}(\vec{S}_{i},\, \vec{S}_{i+1})\! \neq \!
\mathcal{K}(\vec{S}_{i+1},\, \vec{S}_{i})$,  
properties similar to Eqs.~\eqref{completeness} and~\eqref{ortho} hold for the left and right eigenfunctions, but this is not our case.  
Using Eq.~\eqref{completeness} the kernel can be
rewritten as
\begin{eqnarray}
\label{Kernel_TM_eigenf} \mathcal{K}(\vec{S}_{i},\,
\vec{S}_{i+1})=\sum_{m}\lambda _{m}\, \psi_{m}(\vec{S}_{i})\psi
_{m}(\vec{S}_{i+1})\,.
\end{eqnarray}
Combining Eqs.~\eqref{TM_kernel_trace},~\eqref{Kernel_TM_eigenf} and \eqref{ortho} we get 
\begin{eqnarray}
\label{Prtition_TM_all}
 \mathcal{Z}=\sum_{m}\lambda_{m}^{N}\, .
\end{eqnarray}
The eigenvalues $\lambda_{m}$ are all real and positive, as the
kernel operator~\eqref{TM_kernel} is a positive defined function of
$\vec{S}_{i}$ and $\vec{S}_{i+1}$. For symmetric kernels, 
$\mathcal{K}(\vec{S}_{i},\, \vec{S}_{i+1})\!=\!
\mathcal{K}(\vec{S}_{i+1},\, \vec{S}_{i})$, the reality of the
eigenvalues descends from the analogous of the spectral theorem for
real symmetric matrices.  
Moreover, it is possible to show that the spectrum of~\eqref{TM_integral_eq} is upper
bounded so that the eigenvalues $\lambda_{m}$ can be ordered from the
largest to the smallest one:
\[ \lambda_{0}>\lambda_{1}>\lambda_{2}>\ldots \]
In the thermodynamic  limit the asymptotic behavior of the
partition function \eqref{Prtition_TM_all} is dominated by the
largest eigenvalue $\lambda_{0}$, yielding
\begin{eqnarray}
\label{Prtition_TM_asym} \mathcal{Z}\underset{N\rightarrow
\infty}{\sim} \lambda_{0}^{N}\, .
\end{eqnarray}
Once the largest eigenvalue $\lambda_{0}$ is known, the free energy and its derivative can be computed from the relation 
$ F=-T\, \ln \mathcal{Z}$.
%
%%%%%%%%%%%%%%%%%%%%%%%%%%%%%%%%%%%%%%%%%%
\subsection{Pair-spin correlations}
Here we recall~\cite{Tannous,Blume75PRB} how pair-spin correlations 
can be evaluated by means of the eigenvalues and the eigenfunctions defined in 
Eq.~\eqref{TM_integral_eq}. 
Consider the $\mu$  component of the spin located at site $i$ and the $\nu$ component of the spin located at the site
$(i+r)$, with $\mu,\,\nu\!=\!x,y,z$. The average we have to evaluate is
\begin{eqnarray}
&\left\langle S^{\mu }_{i}S^{\nu}_{i+r}\right\rangle\nonumber
=\frac{1}{\mathcal{Z}}\int d\Omega_{1}\int d\Omega_{2}
\ldots \int
\mathcal{K}(\vec{S}_{1},\, \vec{S}_{2}) \\   
&\ldots \,\mathcal{K}(\vec{S}_{i-1},\, \vec{S}_{i})\, S^{\mu
}_{i}\,\mathcal{K}(\vec{S}_{i},\, \vec{S}_{i+1}) \ldots \\   \nonumber
&\mathcal{K}(\vec{S}_{i+r-1},\, \vec{S}_{i+r})\, S^{\nu	
}_{i+r}\,\mathcal{K}(\vec{S}_{i+r},\, \vec{S}_{i+r+1}) \ldots \,
\mathcal{K}(\vec{S}_{N},\, \vec{S}_{1})d\Omega_{N}\,.
\end{eqnarray}
Following the procedure of the previous section  we obtain
\begin{eqnarray}
&\left\langle S^{\mu }_{i}S^{\nu}_{i+r}\right\rangle
=\frac{1}{\mathcal{Z}}\sum_{m_{1},m_{2}
\ldots ,m_{N}}
\lambda_{m_1}\lambda_{m_2}\\ \nonumber
&
\ldots \,\lambda_{m_N}
\delta_{m_{1},m_{2}}\delta_{m_{2},m_{3}} \\ \nonumber
&
\ldots \,
\delta_{m_{i-2},m_{i-1}} 
\int \psi_{m_{i-1}}(\vec{S}_{i})\,S^{\mu }_{i}\,
\psi_{m_{i}}(\vec{S}_{i})d\Omega_{i}\,
\delta_{m_{i},m_{i+1}} \\ \nonumber
&
\ldots \, \delta_{m_{i+r-2},m_{i+r-1}}
\int \psi_{m_{i+r-1}}(\vec{S}_{i+r})\,S^{\nu }_{i+r}\,
 \psi_{m_{i+r}}(\vec{S}_{i+r})d\Omega_{i+r}\\ \nonumber
&
\delta_{m_{i+r},m_{i+r+1}}
\ldots \,\delta_{m_{N-1},m_{N}}\delta_{m_{N},m_{1}} \,.
\end{eqnarray}
Considering all the repeated indices in the Kronecker symbols,
we have
\begin{eqnarray}
\label{correl_funct_TM_before_TD_limit} 
&\left\langle S^{\mu}_{i}S^{\nu}_{i+r}\right\rangle    		\nonumber
\!=\!\frac{1}{\mathcal{Z}}\sum_{m_{i},m_{i+r}} \lambda_{m_{i+r}}^{N-r}
\int \psi_{m_{i+r}}(\vec{S}_{i})\,S^{\mu }_{i}\,
\psi_{m_{i}}(\vec{S}_{i})d\Omega_{i} 
\\ 
&
\lambda_{m_{i}}^{r} \int \psi_{m_{i}}(\vec{S}_{i+r})\,S^{\nu}_{i+r}\,
 \psi_{m_{i+r}}(\vec{S}_{i+r})d\Omega_{i+r}\,.
\end{eqnarray}
If we substitute to  $\mathcal{Z}$ its asymptotic expansion, Eq.~\eqref{Prtition_TM_asym}, we need to evaluate the 
products 
\begin{eqnarray}
\Big(\frac{\lambda_{m_{i+r}}}{\lambda_{0}}\Big)^N\times
\Big(\frac{\lambda_{m_{i}}}{\lambda_{m_{i+r}}}\Big)^r 
\end{eqnarray}
for $N\!\rightarrow\!\infty$. It is straightforward to conclude that only the terms for which 
$m_{i+r}=0$ will not vanish. Thus, pair-spin 
correlations are given by 
\begin{eqnarray}
\label{correl_funct_TM} 
\left\langle S^{\mu }_{i}S^{\nu}_{i+r}\right\rangle  		\nonumber
&=\sum_{m_{i}}
\Big(\frac{\lambda_{m_{i}}}{\lambda_{0}}\Big)^r \int
\psi_{0}(\vec{S}_{i})\,S^{\mu }_{i}\,
 \psi_{m_{i}}(\vec{S}_{i})d\Omega_{i}\\ 
& \int \psi_{m_{i}}(\vec{S}_{i+r})\,S^{\nu }_{i+r}\,
 \psi_{0}(\vec{S}_{i+r})d\Omega_{i+r}\,.
\end{eqnarray}
In the previous formula, $i$ is a dummy index but the result
obviously depends on the separation between the two considered spins, 
$r$. Note that not only the $\lambda_0$
$\psi_{0}$ but \textit{all} the eigenvalues and eigenfunctions
enter Eq.~\eqref{correl_funct_TM}. Eq.~\eqref{correl_funct_TM} can be 
further simplified when the 
$\mu$-$\mu$ correlation function is considered
\begin{eqnarray}
\label{correl_funct_TM_1} 
&\left\langle S^{\mu }_{i}S^{\mu}_{i+r}\right\rangle			\nonumber
=\sum_{m_{i}} \Big(\frac{\lambda_{m_{i}}}{\lambda_{0}}\Big)^r
\Big|\int \psi_{0}(\vec{S}_{i})\,S^{\mu }_{i}\,
 \psi_{m_{i}}(\vec{S}_{i})d\Omega_{i}\Big|^2\,.\\ 
&
\end{eqnarray}
%%%%%%%%%%%%%%%%%%%%%%%%%%%%%%%%%%%%%%%%%%%%%%%%%%%%%%%%%%%%%%%%%%%%%%%%%%%%%%%%%%%
\subsection{Discretization of the unitary sphere} 
The eigenvalue problem defined in Eq.~\eqref{TM_integral_eq} 
can be mapped into a linear algebra problem by sampling the  unitary sphere with a finite number of points. 
Given a generic function of two angles  $\theta $ and  $\phi$, say $f(\theta ,\, \phi )$, the
integral over the solid angle ($d\Omega=d\phi \sin \theta d\theta$) can be approximated as:
\begin{eqnarray}
 \int f(\phi ,\, \theta )\, d\Omega \simeq
\sum ^{P}_{h=1}w_{h}f(u_{h})
\, ,
\end{eqnarray}
where  $ u_{h}$  represent the \textit{special} points that sample
the unitary sphere, $w_{h}$ are the relative weights and $P$ is
number of points themselves~\cite{Stroud}. After 
discretizing the kernel $\mathcal{K}$ in this way, the eigenvalue problem~\eqref{TM_integral_eq} transforms into the following 
linear algebra problem 
\begin{eqnarray}
\label{FSTM_integral equation_sample} \sum ^{P}_{h=1}w_{h}
\mathcal{K} (u_{l},\, u_{h})\, \psi _{m}(u_{h})&=\lambda _{m}\psi_{m}(u_{l})
\end{eqnarray}
which is usually symmetrized  as
\begin{eqnarray*}
\begin{cases}
K_{l,h}&=\sqrt{w_{l}w_{h}}\, \mathcal{K} (u_{l},\, u_{h})\\
\Psi_{h}^{n}&=\sqrt{w_{h}}\, \psi_{n}(u_{h}) 
\end{cases}
\end{eqnarray*}
to yield 
\begin{eqnarray}
\label{FSTM_LA_eigenvalues_equation} \sum ^{P}_{h=1}K_{l,h}\Psi
_{h}^{n}=\lambda _{n}\Psi _{l}^{n}\,.
\end{eqnarray}
The number of special points, $P$, defines the size of the matrix that has to be diagonalized.
For the calculations reported in this work we have chosen 
$P=72,\,128,\, 200$ (Mc Laren formula
14-th degree~\cite{McLaren} and Gauss spherical product formulae of
15-th and 16-th degree~\cite{Abramowitz}). 
The comparison between the results obtained for different sampling 
allows estimating the accuracy of the calculation at a given temperature.  
We used the routine DSPEV of the
LAPACK package to solve the eigenvalue problem defined in Eq.~\eqref{FSTM_LA_eigenvalues_equation}. 
%
%%%%%%%%%%%%%%%%%%%%%%%%%%%%%%%%%%%%%%%%%%%%%%%%%%%%%%%%%%%%%%
%
\section{Polyakov renormalization\label{Polyakov}}
We rewrite here for convenience the continuum Hamiltonian given in Eq.~\eqref{Heisenberg_Ham_cont}:  
\begin{equation} 
\label{Heisenberg_Ham_cont_Poly}
\mathcal{H}=-JN + \int \left[ \frac{J}{2}|\partial _x \vec{S}|^2
-D \left(S^z(x)\right)^2\right] dx
\end{equation}
and represent $\vec{S}(x)$ as a superposition of tow fields: 
\begin{equation} 
\vec{S}(x) = \vec{n}(x)\sqrt{1-\phi^2(x)} + \vec{\phi}(x)\,.
\end{equation} 
As mentioned in the main text, requiring that $|\vec{n}(x)|\!=\!1$ one necessarily has $\vec{n}(x)\cdot \vec{\phi}(x)\!=\!0$ so that
$\vec{\phi}(x)$ can be expressed on a basis orthonormal to $\vec{n}(x)$: 
\begin{equation} 
 \vec{\phi}(x)= \sum_a \phi_a \vec{e}_a
\quad\quad\text{with  } a=1,2
\end{equation}  
with $|\vec{e}_a(x)|=1$~\cite{footnote_Polyakov}. 
Moreover, the fact that $|\vec{n}(x)|\!=\!1$ implies that 
$\partial _x \vec{n}(x)$ is orthogonal to $\vec{n}(x)$ itself, thus
\begin{equation} 
 \partial _x \vec{n}(x) = \sum_a c_a \vec{e}_a\,.
\end{equation}  
The gradient term in Hamiltonian~\eqref{Heisenberg_Ham_cont_Poly} reads
\begin{eqnarray} 
\begin{split}
|\partial _x \vec{S}|^2 &= |  \sqrt{1-\phi^2}\partial _x \vec{n} + \partial _x\sqrt{1-\phi^2} \vec{n}   \\
&+\sum_a \partial _x \phi_a \vec{e}_a +  \sum_a \phi_a \partial _x \vec{e}_a |^2 \,.
\end{split}
\end{eqnarray} 
Exploiting the orthogonality of the basis $(\vec{n},\vec{e}_a)$, the derivatives $\partial _x \vec{e}_a$ can be written as 
\begin{equation} 
\partial _x \vec{e}_a = -c_a \vec{n}+ f_{ab}\vec{e}_b
\end{equation} 
where $f_{ab}$ is an antisymmetric two-by-two tensor whose components can be made negligible with a proper choice of the 
reference frame $(\vec{e}_1, \vec{e}_2)$.  
Finally, one gets 
\begin{equation} 
\begin{split} 
\label{grad_total_Poly}
|\partial _x \vec{S}|^2&=\left(1-\phi^2\right) \left(\partial _x\vec{n}\right)^2 + \sum_{a} \left( \partial_x \phi_a \right)^2+ \sum_{ab} c_a c_b\phi_a\phi_b \\
&+2 \sqrt{1-\phi^2}\sum_{a} \partial_x \phi_a c_a-2\partial _x\sqrt{1-\phi^2}\sum_a c_a \phi_a\\
&+\left( \partial _x\sqrt{1-\phi^2}\right)^2  \,.
\end{split} 
\end{equation}
The two terms at the second line of Eq.~\eqref{grad_total_Poly} vanish after spatial integration; while the last term (third line) 
only contains powers of $\phi_a$ higher than the square after the derivation (thus we neglect it).
$\sum_{a} \left( \partial_x \phi_a \right)^2$ represents the ``kinetic'' term of the field $\vec{\phi}(x)$. Our goal is to retain all the other 
quadratic terms in the field  $\vec{\phi}(x)$ and use them to perform thermal averages. 
After thermal averaging, $\langle \dots \rangle$, we will  have   
\begin{equation} 
\langle \phi_a\phi_b \rangle = \delta_{ab} \langle \phi_a^2 \rangle 
\end{equation}
which allows rewriting 
\begin{equation} 
\langle\sum_{ab} c_a c_b\phi_a\phi_b  \rangle =  \langle \phi_a^2 \rangle \sum_{a} c_a^2 = \langle \phi_a^2 \rangle\left(\partial _x\vec{n}\right)^2\,. 
\end{equation}
Besides this, one has
\begin{equation} 
\langle \phi^2 \rangle = \langle \phi_1^2 + \phi_2^2 \rangle =   2\langle \phi_a^2 \rangle \,.
\end{equation}
Thus, the Polyakov's result is readily recovered 
\begin{equation} 
\label{Gradient_ren_Poly}
\langle|\partial _x \vec{S}|^2\rangle=\left[1-\langle \phi_a^2 \rangle \right]  \left(\partial _x\vec{n}\right)^2 
+ \sum_{a} \langle\left( \partial_x \phi_a \right)^2 \rangle \,.
\end{equation}
Let us consider now the anisotropy term in Hamiltonian~\eqref{Heisenberg_Ham_cont_Poly}: 
\begin{equation} 
\begin{split} 
\left(S^z\right)^2 & = \left( n^z \sqrt{1-\phi^2} + \sum_a \phi_a e^z_a  \right)^2 \\
&=\left( n^z\right)^2 \left(1-\phi^2\right) + \sum_{ab} \phi_a \phi_b e^z_a e^z_b + \mathcal{O}\left(\phi_a \right)\,,   
\end{split} 
\end{equation} 
where $\mathcal{O}\left(\phi_a \right)$ stands for terms linear in $ \phi_a$ that vanish after spatial integration; while after thermal averaging one has
\begin{equation} 
\langle\sum_{ab} \phi_a \phi_b e^z_a e^z_b\rangle = \langle \phi_a^2\rangle\sum_{a} (e^z_a)^2
= \langle \phi_a^2\rangle \left[1-\left( n^z\right)^2 \right]\,;
\end{equation} 
In this case, we obtain 
\begin{equation}
\label{S_z_sq_ren_Poly}  
\langle\left(S^z\right)^2 \rangle= \left[1-3\langle \phi_a^2\rangle\right] \left( n^z\right)^2 + \langle \phi_a^2\rangle\,.
\end{equation} 
%%%%%%%%%%%%%%%%%%%%%%%%%%%%%%%%%%%%%%%%%%%%%%%%%%%%%%%%%%%%%%%%%%%%%%%%%%%%%%%

Now we want to derive the quadratic Hamiltonian for the $\vec \phi$ field written 
in Eq.~\eqref{SW_Ham} and used to perform thermal averages. 
We first rewrite the Hamiltonian~\eqref{Heisenberg_Ham_cont_Poly} 
in terms of the fields $\vec{\phi}(x)$ and $\vec{n}(x)$. In doing this, we will 
keep the approximations we made previously: we will neglect the terms of order higher than the second in $\phi_a$ 
as well as what is supposed to vanish after the spatial integration. A delicate point of the latter approximation is the assumption
\begin{equation} 
\phi_a\phi_b  = \delta_{ab} \frac{ \phi_1^2 + \phi_2^2 }{2} 
\end{equation}
based on symmetry reasons. 
The simplified continuum Hamiltonian reads 
\begin{equation} 
\begin{split} 
\label{Heisenberg_Ham_cont_phi_Poly}
\mathcal{H}&=-JN + \frac{J}{2} \int\left[1-\frac{\phi_1^2 + \phi_2^2}{2} \right]|\partial _x \vec{n}|^2 dx \\
&+ \sum_{a}  \frac{J}{2} \int \left( \partial_x \phi_a \right)^2 dx\\
&-D \int\left[\left( n^z\right)^2\left(1-3 \frac{\phi_1^2 + \phi_2^2}{2} \right) +  \frac{\phi_1^2 + \phi_2^2}{2}  \right] dx\\
&=-JN + \frac{J}{2} \int|\partial _x \vec{n}|^2 dx -D\int \left( n^z\right)^2 dx \\
&+ \sum_{a}  \frac{J}{2} \int \left( \partial_x \phi_a \right)^2 dx \\
&+\sum_a \int \left[ \frac{3D}{2} \left(n^z\right)^2-\frac{D}{2}-\frac{J}{2} |\partial _x \vec{n}|^2\right]\phi_a^2 dx\,.
\end{split} 
\end{equation}
%%%%%%%%%%%%%%%%%%%%%%%%%%%%%%%%%%%%%%%%%%%%%%%%%%%%%%%%%%%%%%%%%%%%%%%%%%%%%%%
Within a distance separating two successive DWs we assume 
\begin{equation} 
\begin{cases} 
&\left(n^z\right)^2 \simeq 1 \\
&|\partial _x \vec{n}|^2\simeq 0 \,,
\end{cases} 
\end{equation}
which allows expressing the $\vec{\phi}$-field Hamiltonian as  
\begin{equation} 
\label{SW_Ham_Poly}
\mathcal{H}_{\phi} = \frac{N}{\Lambda}\sum_{a}  \int_{-\Lambda/2}^{\Lambda/2}  \left[ \frac{J}{2} \left( \partial_x \phi_a \right)^2 + D \phi_a^2 \right] dx \, ,
\end{equation}
$\Lambda$ being the average distance between two successive DWs. 
%
%
%%%%%%%%%%%%%%%%%%%%%%%%%%%%%%%%%%%%%%%%%%%%%%%%%%%%%%%%%%%%%%%%%%%%%%%%%%%%%
%
\section{Domain-wall motion and random walk\label{App_DW_traj}}  
\begin{figure}%[ht]
\includegraphics*[width=8cm,angle=0]{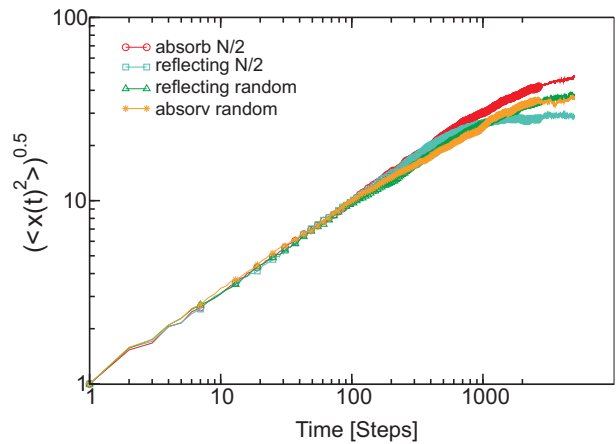}
\caption{\label{fig4}Color online. Square root of the  mean square deviation of a random walker obtained using different 
boundary and initial conditions. }
\end{figure}

In order to check how the choice of boundary conditions could affect DW diffusion, we performed a preliminary analysis of a 
standard one-dimensional random walk constrained to a finite segment. 
Such a geometrical constraint is equivalent to the one experienced by a DW which diffuses in the 
Heisenberg spin chain with o.b.c..   
In a random walk process the relation between the mean square displacement $\langle  x^2 \rangle$ and the time $t$ elapsed during 
the walk is:  
%
%%%%  EQUATION 3 %%%%%%%%%%%%%%%%%%%%%%%%%%%%%%%%%%%%%%%%%%%%%%
\begin{equation}\label{eq3}
\langle x^2 \rangle = \frac{\sigma^2}{\tau_c} t,
\end{equation}
%%%%%%%%%%%%%%%%%%%%%%%%%%%%%%%%%%%%%%%%%%%%%%%%%%%%%%%%%%%%%%%
\noindent where $\sigma$ is the mean square displacement and $\tau_c$ is the mean-time for each step. 
We considered a discrete random walk in a linear system of $N$ sites using different boundary 
and initial conditions.  Since we set the lattice unit $a\!=\!1$ and take $\tau_c \!=\! 1$ then $\sigma^2 \!=\!1$.
In Fig.~\ref{fig4}, we plot the square root of the mean-square displacement in a system of the same size of 
our spin chain, $N\!=\!100$. 
We used both reflecting and absorbing boundary conditions. In addition, we chose for each case two initial conditions: 
in one we started the random walk from the middle of the chain and in the other one at a random position. 
In the four studied cases we observed that any significant deviation of the relation given by Eq.~\eqref{eq3} 
started at $\langle  x^2 \rangle \!>\! 10$. Below this value boundary conditions are not expected to affect  
the analysis of DW trajectories either.  
%
%%%% FIGURA 4 %%%%%%%%%%%%%%%%%%%%%%%%%%%%%%%%%%%%%%%%%%%%%%%%%%%%%%%%%%%%%%%%%%%%%%%%%%%%%%%%%%%%%%%%%%
%%%%%%%%%%%%%%%%%%%%%%%%%%%%%%%%%%%%%%%%%%%%%%%%%%%%%%%%%%%
%%%% FIGURA 6 %%%%%%%%%%%%%%%%%%%%%%%%%%%%%%%%%%%%%%%%%%%
\begin{figure}
\includegraphics*[width=8cm,angle=0]{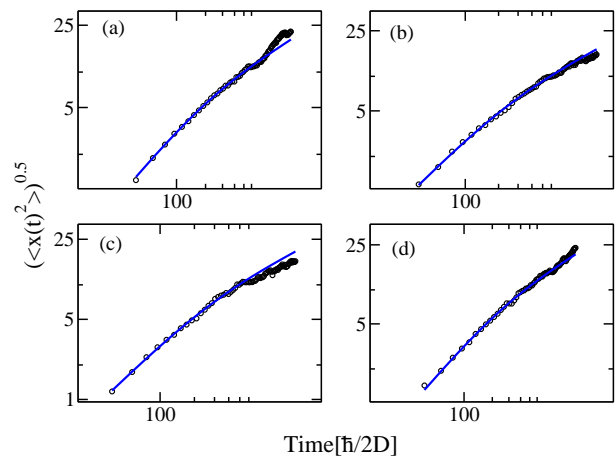}
\caption{\label{fig6} Average dispersion of the DW trajectories at $T/J\!=\!0.14$ for two damping constant and applied
fields In Figs.  (a) and (c) the applied field is $H\!=\!0$ and the damping constant $a\!=\!0.25$ and $a\!=\!4$, respectively. 
In Figs.  (b) and (d) the applied field $H\!=\!0.01$ and the damping constant $a\!=\!0.25$ and $a\!=\!4$, respectively.}
\end{figure}
%%%%%%%%%%%%%%%%%%%%%%%%%%%%%%%%%%%%%%%%%%%%%%%%%%%%%%%%%%%%%%%%%%%%%%%%%%%%%%% FIGURA 8 %%%%%%%%%%%%%%%%%%%%%%%%%%%%%%%%%%%%%%%%%%%%%%%%%%%%%%%%%%%%%%%%%%%%%%%%%%%%%%%%
\begin{figure}
\includegraphics*[width=8cm,angle=0]{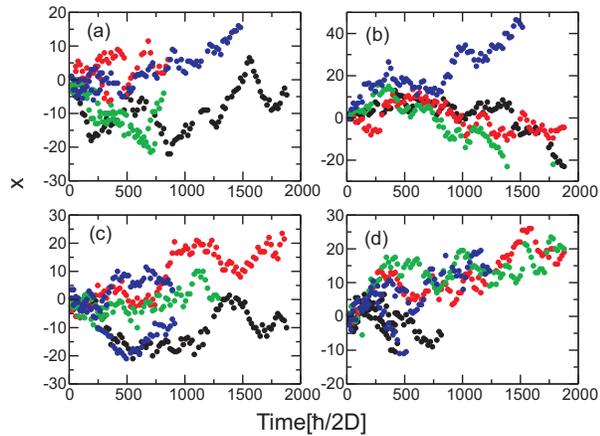}
\caption{\label{fig8} Color online. Domain-wall trajectories at $T/J\!=\!0.14$ for two damping constant and applied
fields. In Figs.  (a) and (c) the applied field is $H\!=\!0$ and the damping constant $a\!=\!0.25$ and $a\!=\!4$, respectively. 
In Figs.  (b) and (d) the applied field $H\!=\!0.01$ and the damping constant $a\!=\!0.25$ and $a\!=\!4$, respectively.}
\end{figure}
%%%%%%%%%%%%%%%%%%%%%%%%%%%%%%%%%%%%%%%%%%%%%%%%%%%%%%%%%%%%%%%%%%%%%%%%%%%

For what concerns the spin chain, we began by studying the diffusion of DWs which were naturally present in the system at the chosen temperature 
(if $\xi\!<\!N$).  
The  setup was the same as the one used to obtain the curves of Fig.~\ref{fig2}. 
We chose three representative temperatures, $T/J \!= \! 0.22$, $T/J \! = \! 0.18$, and $T/J \! = \! 0.14$ 
using two values for the damping constant:  $a \! = \! 0.25$ and $ a \! = \! 4$.
Fig.~\ref{fig6} shows the mean-square DW displacement when the system was equilibrated at  $T/J\!=\!0.14$, for $D/J\!=\!0.2$. 
The data reported in the two figures on the left-hand-side correspond to zero applied field, 
whereas for the two figures on the right-hand-side the field was $H/J\!=\!0.01$. 
Instead, figures in the first (Figs.~\ref{fig6} (a) and (b)) and in the second  raw (Figs.~\ref{fig6} (c) and (d)) 
correspond to the two different values of the damping constant. 
For all the curves, the diffusion relation $\langle x^2\rangle \propto t$ is not obeyed at short times (see Eq.~\eqref{eq3}).  
Thus we fitted those trajectories with a more general expression which describes a random walk with correlated steps~\cite{Taylor20PLMS}: 
%
%%%%  EQUATION 4 %%%%%%%%%%%%%%%%%%%%%%%%%%%%%%%%%%%%%%%%%%%%%%%%%%%%%%%%%%%
\begin{equation}\label{eq4_app}  
\langle x^2 \rangle  = \sigma^2 \left( \frac{t}{\tau_c} + \frac{e^{-\frac{2t}{\tau_c}} -1}{2} \right) \, . 
\end{equation}
%%%%%%%%%%%%%%%%%%%%%%%%%%%%%%%%%%%%%%%%%%%%%%%%%%%%%%%%%%%%%%%%%%%%%%%%%%%%%
%
\noindent the meaning of the parameters Eq.~\eqref{eq4_app} is explained in the main text in Sect.~\ref{sec_V}.  

Fig.~\ref{fig8} displays some individual trajectories that were used to produce  the
curves in Fig.~\ref{fig6}. Some correlation emerges for  small displacements, of the
order of the DW width, which confirms the occurrence of a ballistic regime for \textit{short} times. 
At very short time scales there is an uncertainty in the DW  
position of the order of a lattice parameter. In fact, fast fluctuations of the DW structure 
produce an offset in the initial position of the diffusing DW, intrinsic to the method used to
detect such a position. 

\bibliographystyle{apsrev4-1}
\bibliography{Billoni}

%Merlin.mbs v4.21 2009-07-09.
\begin{thebibliography}{10}%
\makeatletter
\providecommand \@ifxundefined [1]{%
 \ifx #1\undefined \expandafter \@firstoftwo
 \else \expandafter \@secondoftwo
\fi
}%
\providecommand \@ifnum [1]{%
 \ifnum #1\expandafter \@firstoftwo
 \else \expandafter \@secondoftwo
\fi
}%
\providecommand \enquote [1]{``#1''}%
\providecommand \bibnamefont  [1]{#1}%
\providecommand \bibfnamefont [1]{#1}%
\providecommand \citenamefont [1]{#1}%
\providecommand\href[0]{\@sanitize\@href}%
\providecommand\@href[1]{\endgroup\@@startlink{#1}\endgroup\@@href}%
\providecommand\@@href[1]{#1\@@endlink}%
\providecommand \@sanitize [0]{\begingroup\catcode`\&12\catcode`\#12\relax}%
\@ifxundefined \pdfoutput {\@firstoftwo}{%
 \@ifnum{\z@=\pdfoutput}{\@firstoftwo}{\@secondoftwo}%
}{%
 \providecommand\@@startlink[1]{\leavevmode\special{html:<a href="#1">}}%
 \providecommand\@@endlink[0]{\special{html:</a>}}%
}{%
 \providecommand\@@startlink[1]{%
  \leavevmode
  \pdfstartlink
   attr{/Border[0 0 1 ]/H/I/C[0 1 1]}%
   user{/Subtype/Link/A<</Type/Action/S/URI/URI(#1)>>}%
  \relax
 }%
 \providecommand\@@endlink[0]{\pdfendlink}%
}%
\providecommand \url  [0]{\begingroup\@sanitize \@url }%
\providecommand \@url [1]{\endgroup\@href {#1}{\urlprefix}}%
\providecommand \urlprefix [0]{URL }%
\providecommand \Eprint[0]{\href }%
\@ifxundefined \urlstyle {%
  \providecommand \doi [1]{doi:\discretionary{}{}{}#1}%
}{%
  \providecommand \doi [0]{doi:\discretionary{}{}{}\begingroup
  \urlstyle{rm}\Url }%
}%
\providecommand \doibase [0]{http://dx.doi.org/}%
\providecommand \Doi[1]{\href{\doibase#1}}%
\providecommand \bibAnnote [3]{%
  \BibitemShut{#1}%
  \begin{quotation}\noindent
    \textsc{Key:}\ #2\\\textsc{Annotation:}\ #3%
  \end{quotation}%
}%
\providecommand \bibAnnoteFile [2]{%
  \IfFileExists{#2}{\bibAnnote {#1} {#2} {\input{#2}}}{}%
}%
\providecommand \typeout [0]{\immediate \write \m@ne }%
\providecommand \selectlanguage [0]{\@gobble}%
\providecommand \bibinfo [0]{\@secondoftwo}%
\providecommand \bibfield [0]{\@secondoftwo}%
\providecommand \translation [1]{[#1]}%
\providecommand \BibitemOpen[0]{}%
\providecommand \bibitemStop [0]{}%
\providecommand \bibitemNoStop [0]{.\EOS\space}%
\providecommand \EOS [0]{\spacefactor3000\relax}%
\providecommand \BibitemShut [1]{\csname bibitem#1\endcsname}%
%</preamble>
\bibitem{Slonczewski_JMMM_96}%
  \BibitemOpen
  \bibfield{author}{%
  \bibinfo {author} {\bibfnamefont{J.~C.}\ \bibnamefont{Slonczewski}},\ }%
  \bibfield{journal}{%
  \Doi{doi:10.1016/0304-8853(96)00062-5}{\bibinfo {journal} {J. Magn. Magn.
  Mater.}}\ }%
  \textbf{\bibinfo {volume} {159}},\ \bibinfo {pages} {L1} (\bibinfo {year}
  {1996})%
  \bibAnnoteFile{NoStop}{Slonczewski_JMMM_96}%
\bibitem{Parkin11042008}%
  \BibitemOpen
  \bibfield{author}{%
  \bibinfo {author} {\bibfnamefont{S.~S.~P.}\ \bibnamefont{Parkin}}, \bibinfo
  {author} {\bibfnamefont{M.}~\bibnamefont{Hayashi}},\ and\ \bibinfo {author}
  {\bibfnamefont{L.}~\bibnamefont{Thomas}},\ }%
  \bibfield{journal}{%
  \Doi{10.1126/science.1145799}{\bibinfo {journal} {Science}}\ }%
  \textbf{\bibinfo {volume} {320}},\ \bibinfo {pages} {190} (\bibinfo {year}
  {2008})%
  \bibAnnoteFile{NoStop}{Parkin11042008}%
\bibitem{Caneschi01ACIE}%
  \BibitemOpen
  \bibfield{author}{%
  \bibinfo {author} {\bibfnamefont{A.}~\bibnamefont{Caneschi}}, \bibinfo
  {author} {\bibfnamefont{D.}~\bibnamefont{Gatteschi}}, \bibinfo {author}
  {\bibfnamefont{N.}~\bibnamefont{Lalioti}}, \bibinfo {author}
  {\bibfnamefont{C.}~\bibnamefont{Sangregorio}}, \bibinfo {author}
  {\bibfnamefont{R.}~\bibnamefont{Sessoli}}, \bibinfo {author}
  {\bibfnamefont{G.}~\bibnamefont{Venturi}}, \bibinfo {author}
  {\bibfnamefont{A.}~\bibnamefont{Vindigni}}, \bibinfo {author}
  {\bibfnamefont{A.}~\bibnamefont{Rettori}}, \bibinfo {author}
  {\bibfnamefont{M.~G.}\ \bibnamefont{Pini}},\ and\ \bibinfo {author}
  {\bibfnamefont{M.~A.}\ \bibnamefont{Novak}},\ }%
  \bibfield{journal}{%
  \bibinfo {journal} {Angew. Chem. Int. Ed.}\ }%
  \textbf{\bibinfo {volume} {40}},\ \bibinfo {pages} {1760} (\bibinfo {year}
  {2001})%
  \bibAnnoteFile{NoStop}{Caneschi01ACIE}%
\bibitem{Caneschi02EPL}%
  \BibitemOpen
  \bibfield{author}{%
  \bibinfo {author} {\bibfnamefont{A.}~\bibnamefont{Caneschi}}, \bibinfo
  {author} {\bibfnamefont{D.}~\bibnamefont{Gatteschi}}, \bibinfo {author}
  {\bibfnamefont{N.}~\bibnamefont{Lalioti}}, \bibinfo {author}
  {\bibfnamefont{C.}~\bibnamefont{Sangregorio}}, \bibinfo {author}
  {\bibfnamefont{R.}~\bibnamefont{Sessoli}}, \bibinfo {author}
  {\bibfnamefont{G.}~\bibnamefont{Venturi}}, \bibinfo {author}
  {\bibfnamefont{A.}~\bibnamefont{Vindigni}}, \bibinfo {author}
  {\bibfnamefont{A.}~\bibnamefont{Rettori}}, \bibinfo {author}
  {\bibfnamefont{M.~G.}\ \bibnamefont{Pini}},\ and\ \bibinfo {author}
  {\bibfnamefont{M.~A.}\ \bibnamefont{Novak}},\ }%
  \bibfield{journal}{%
  \bibinfo {journal} {Europhys. Lett.}\ }%
  \textbf{\bibinfo {volume} {58}},\ \bibinfo {pages} {771} (\bibinfo {year}
  {2002})%
  \bibAnnoteFile{NoStop}{Caneschi02EPL}%
\bibitem{Clerac02JACS}%
  \BibitemOpen
  \bibfield{author}{%
  \bibinfo {author} {\bibfnamefont{R.}~\bibnamefont{Cl\'erac}}, \bibinfo
  {author} {\bibfnamefont{H.}~\bibnamefont{Miyasaka}}, \bibinfo {author}
  {\bibfnamefont{M.}~\bibnamefont{Yamashita}},\ and\ \bibinfo {author}
  {\bibfnamefont{C.}~\bibnamefont{Coulon}},\ }%
  \bibfield{journal}{%
  \Doi{10.1021/ja0203115}{\bibinfo {journal} {J. Am. Chem. Soc.}}\ }%
  \textbf{\bibinfo {volume} {124}},\ \bibinfo {pages} {12837} (\bibinfo {month}
  {Oct.}\ \bibinfo {year} {2002})%
  \bibAnnoteFile{NoStop}{Clerac02JACS}%
\bibitem{Lescouzec03ACIE}%
  \BibitemOpen
  \bibfield{author}{%
  \bibinfo {author} {\bibfnamefont{R.}~\bibnamefont{Lescou\"ezec}}, \bibinfo
  {author} {\bibfnamefont{J.}~\bibnamefont{Vaissermann}}, \bibinfo {author}
  {\bibfnamefont{C.}~\bibnamefont{Ruiz-P\'erez}}, \bibinfo {author}
  {\bibfnamefont{F.}~\bibnamefont{Lloret}}, \bibinfo {author}
  {\bibfnamefont{R.}~\bibnamefont{Carrasco}}, \bibinfo {author}
  {\bibfnamefont{M.}~\bibnamefont{Julve}}, \bibinfo {author}
  {\bibfnamefont{M.}~\bibnamefont{Verdaguer}}, \bibinfo {author}
  {\bibfnamefont{Y.}~\bibnamefont{Dromzee}}, \bibinfo {author}
  {\bibfnamefont{D.}~\bibnamefont{Gatteschi}},\ and\ \bibinfo {author}
  {\bibfnamefont{W.}~\bibnamefont{Wernsdorfer}},\ }%
  \bibfield{journal}{%
  \Doi{10.1002/anie.200250243}{\bibinfo {journal} {Angew. Chem. Int. Ed.}}\ }%
  \textbf{\bibinfo {volume} {42}},\ \bibinfo {pages} {1427} (\bibinfo {year}
  {2003})%
  \bibAnnoteFile{NoStop}{Lescouzec03ACIE}%
\bibitem{Gatteschi-Sessoli_rev_SMM_03}%
  \BibitemOpen
  \bibfield{author}{%
  \bibinfo {author} {\bibfnamefont{D.}~\bibnamefont{Gatteschi}}\ and\ \bibinfo
  {author} {\bibfnamefont{R.}~\bibnamefont{Sessoli}},\ }%
  \bibfield{journal}{%
  \Doi{10.1002/anie.200390099}{\bibinfo {journal} {Angew. Chem. Int. Ed.}}\ }%
  \textbf{\bibinfo {volume} {42}},\ \bibinfo {pages} {268} (\bibinfo {year}
  {2003})%
  \bibAnnoteFile{NoStop}{Gatteschi-Sessoli_rev_SMM_03}%
\bibitem{Miyasaka_review}%
  \BibitemOpen
  \bibfield{author}{%
  \bibinfo {author} {\bibfnamefont{M.~Y.}\ \bibnamefont{Hitoshi~Miyasaka},
  \bibfnamefont{Miguel~Julve}}\ and\ \bibinfo {author}
  {\bibfnamefont{R.}~\bibnamefont{Cl\'erac}},\ }%
  \bibfield{journal}{%
  \Doi{10.1021/ic801990g}{\bibinfo {journal} {Inorg. Chem.}}\ }%
  \textbf{\bibinfo {volume} {48}},\ \bibinfo {pages} {3420} (\bibinfo {year}
  {2009})%
  \bibAnnoteFile{NoStop}{Miyasaka_review}%
\bibitem{Coulon06Springer}%
  \BibitemOpen
  \bibfield{author}{%
  \bibinfo {author} {\bibfnamefont{C.}~\bibnamefont{Coulon}}, \bibinfo {author}
  {\bibfnamefont{H.}~\bibnamefont{Miyasaka}},\ and\ \bibinfo {author}
  {\bibfnamefont{R.}~\bibnamefont{Cl\'erac}},\ }%
  \bibfield{journal}{%
  \bibinfo {journal} {Struct. Bond.}\ }%
  \textbf{\bibinfo {volume} {122}},\ \bibinfo {pages} {163} (\bibinfo {year}
  {2006})%
  \bibAnnoteFile{NoStop}{Coulon06Springer}%
\bibitem{Bogani_JMC_08}%
  \BibitemOpen
  \bibfield{author}{%
  \bibinfo {author} {\bibfnamefont{L.}~\bibnamefont{Bogani}}, \bibinfo {author}
  {\bibfnamefont{A.}~\bibnamefont{Vindigni}}, \bibinfo {author}
  {\bibfnamefont{R.}~\bibnamefont{Sessoli}},\ and\ \bibinfo {author}
  {\bibfnamefont{D.}~\bibnamefont{Gatteschi}},\ }%
  \bibfield{journal}{%
  \Doi{http://dx.doi.org/10.1039/b807824f}{\bibinfo {journal} {J. Mater.
  Chem.}}\ }%
  \textbf{\bibinfo {volume} {18}},\ \bibinfo {pages} {4750} (\bibinfo {year}
  {2008})%
  \bibAnnoteFile{NoStop}{Bogani_JMC_08}%
\bibitem{Tobochnik}%
  \BibitemOpen
  \bibfield{author}{%
  \bibinfo {author} {\bibfnamefont{R.}~\bibnamefont{Cordery}}, \bibinfo
  {author} {\bibfnamefont{S.}~\bibnamefont{Sarker}},\ and\ \bibinfo {author}
  {\bibfnamefont{J.}~\bibnamefont{Tobochnik}},\ }%
  \bibfield{journal}{%
  \Doi{10.1103/PhysRevB.24.5402}{\bibinfo {journal} {Phys. Rev. B (R)}}\ }%
  \textbf{\bibinfo {volume} {24}},\ \bibinfo {pages} {5402} (\bibinfo {year}
  {1981})%
  \bibAnnoteFile{NoStop}{Tobochnik}%
\bibitem{Fisher63AJP}%
  \BibitemOpen
  \bibfield{author}{%
  \bibinfo {author} {\bibfnamefont{M.~E.}\ \bibnamefont{Fisher}},\ }%
  \bibfield{journal}{%
  \Doi{10.1119/1.1970340}{\bibinfo {journal} {Am. J. Phys.}}\ }%
  \textbf{\bibinfo {volume} {32}},\ \bibinfo {pages} {343} (\bibinfo {year}
  {1964})%
  \bibAnnoteFile{NoStop}{Fisher63AJP}%
\bibitem{Blume75PRB}%
  \BibitemOpen
  \bibfield{author}{%
  \bibinfo {author} {\bibfnamefont{M.}~\bibnamefont{Blume}}, \bibinfo {author}
  {\bibfnamefont{P.}~\bibnamefont{Heller}},\ and\ \bibinfo {author}
  {\bibfnamefont{N.~A.}\ \bibnamefont{Lurie}},\ }%
  \bibfield{journal}{%
  \Doi{10.1103/PhysRevB.11.4483}{\bibinfo {journal} {Phys. Rev. B}}\ }%
  \textbf{\bibinfo {volume} {11}},\ \bibinfo {pages} {4483} (\bibinfo {month}
  {Jun}\ \bibinfo {year} {1975})%
  \bibAnnoteFile{NoStop}{Blume75PRB}%
\bibitem{Vindigni06APA}%
  \BibitemOpen
  \bibfield{author}{%
  \bibinfo {author} {\bibfnamefont{A.}~\bibnamefont{Vindigni}}, \bibinfo
  {author} {\bibfnamefont{A.}~\bibnamefont{Rettori}}, \bibinfo {author}
  {\bibfnamefont{M.}~\bibnamefont{Pini}}, \bibinfo {author}
  {\bibfnamefont{C.}~\bibnamefont{Carbone}},\ and\ \bibinfo {author}
  {\bibfnamefont{P.}~\bibnamefont{Gambardella}},\ }%
  \bibfield{journal}{%
  \Doi{10.1007/s00339-005-3364-4}{\bibinfo {journal} {Appl. Phys. A}}\ }%
  \textbf{\bibinfo {volume} {82}},\ \bibinfo {pages} {385} (\bibinfo {month}
  {Feb.}\ \bibinfo {year} {2006})%
  \bibAnnoteFile{NoStop}{Vindigni06APA}%
\bibitem{Nowak00PRL}%
  \BibitemOpen
  \bibfield{author}{%
  \bibinfo {author} {\bibfnamefont{U.}~\bibnamefont{Nowak}}, \bibinfo {author}
  {\bibfnamefont{R.~W.}\ \bibnamefont{Chantrell}},\ and\ \bibinfo {author}
  {\bibfnamefont{E.~C.}\ \bibnamefont{Kennedy}},\ }%
  \bibfield{journal}{%
  \bibinfo {journal} {Phys. Rev. Lett.}\ }%
  \textbf{\bibinfo {volume} {84}},\ \bibinfo {pages} {163} (\bibinfo {year}
  {2000})%
  \bibAnnoteFile{NoStop}{Nowak00PRL}%
\bibitem{Cheng06PRL}%
  \BibitemOpen
  \bibfield{author}{%
  \bibinfo {author} {\bibfnamefont{W.~T.}\ \bibnamefont{Cheng}}, \bibinfo
  {author} {\bibfnamefont{M.~B.~A.}\ \bibnamefont{Jalil}}, \bibinfo {author}
  {\bibfnamefont{H.~K.}\ \bibnamefont{Lee}},\ and\ \bibinfo {author}
  {\bibfnamefont{Y.}~\bibnamefont{Okabe}},\ }%
  \bibfield{journal}{%
  \bibinfo {journal} {Phys. Rev. Lett.}\ }%
  \textbf{\bibinfo {volume} {96}},\ \bibinfo {pages} {067208} (\bibinfo {year}
  {2006})%
  \bibAnnoteFile{NoStop}{Cheng06PRL}%
\bibitem{Billoni07JMMM}%
  \BibitemOpen
  \bibfield{author}{%
  \bibinfo {author} {\bibfnamefont{O.~V.}\ \bibnamefont{Billoni}}\ and\
  \bibinfo {author} {\bibfnamefont{D.~A.}\ \bibnamefont{Stariolo}},\ }%
  \bibfield{journal}{%
  \bibinfo {journal} {J. Magn. Magn. Mater.}\ }%
  \textbf{\bibinfo {volume} {316}},\ \bibinfo {pages} {49} (\bibinfo {year}
  {2007})%
  \bibAnnoteFile{NoStop}{Billoni07JMMM}%
\bibitem{Hinzke00PRB}%
  \BibitemOpen
  \bibfield{author}{%
  \bibinfo {author} {\bibfnamefont{D.}~\bibnamefont{Hinzke}}\ and\ \bibinfo
  {author} {\bibfnamefont{U.}~\bibnamefont{Nowak}},\ }%
  \bibfield{journal}{%
  \Doi{10.1103/PhysRevB.61.6734}{\bibinfo {journal} {Phys. Rev. B}}\ }%
  \textbf{\bibinfo {volume} {61}},\ \bibinfo {pages} {6734} (\bibinfo {month}
  {Mar}\ \bibinfo {year} {2000})%
  \bibAnnoteFile{NoStop}{Hinzke00PRB}%
\bibitem{Hinzke01Proc}%
  \BibitemOpen
  \bibfield{author}{%
  \bibinfo {author} {\bibfnamefont{D.}~\bibnamefont{Hinzke}}, \bibinfo {author}
  {\bibfnamefont{U.}~\bibnamefont{Nowak}},\ and\ \bibinfo {author}
  {\bibfnamefont{D.}~\bibnamefont{Usadel}},\ }%
  \bibfield{journal}{%
  \bibinfo {journal} {Proc. structure and dynamics of heterogenous systems,
  Duisburg, Germany, 24 - 26 February 1999},\ \bibinfo {pages} {331}}%
   (\bibinfo {year} {2001})%
  \bibAnnoteFile{NoStop}{Hinzke01Proc}%
\bibitem{Glauber63JMathP}%
  \BibitemOpen
  \bibfield{author}{%
  \bibinfo {author} {\bibfnamefont{R.~J.}\ \bibnamefont{Glauber}},\ }%
  \bibfield{journal}{%
  \Doi{10.1063/1.1703954}{\bibinfo {journal} {J. Math. Phys.}}\ }%
  \textbf{\bibinfo {volume} {4}},\ \bibinfo {pages} {294} (\bibinfo {month}
  {Feb.}\ \bibinfo {year} {1963})%
  \bibAnnoteFile{NoStop}{Glauber63JMathP}%
\bibitem{Barma_J_Stat_Phys_80}%
  \BibitemOpen
  \bibfield{author}{%
  \bibinfo {author} {\bibfnamefont{D.}~\bibnamefont{Dhar}}\ and\ \bibinfo
  {author} {\bibfnamefont{M.}~\bibnamefont{Barma}},\ }%
  \bibfield{journal}{%
  \Doi{10.1007/BF01008051}{\bibinfo {journal} {J. Stat. Phys.}}\ }%
  \textbf{\bibinfo {volume} {22}},\ \bibinfo {pages} {259} (\bibinfo {year}
  {1980})%
  \bibAnnoteFile{NoStop}{Barma_J_Stat_Phys_80}%
\bibitem{Luscombe_PRE_96}%
  \BibitemOpen
  \bibfield{author}{%
  \bibinfo {author} {\bibfnamefont{J.~H.}\ \bibnamefont{Luscombe}}, \bibinfo
  {author} {\bibfnamefont{M.}~\bibnamefont{Luban}},\ and\ \bibinfo {author}
  {\bibfnamefont{J.~P.}\ \bibnamefont{Reynolds}},\ }%
  \bibfield{journal}{%
  \Doi{10.1103/PhysRevE.53.5852}{\bibinfo {journal} {Phys. Rev. E}}\ }%
  \textbf{\bibinfo {volume} {53}},\ \bibinfo {pages} {5852} (\bibinfo {year}
  {1996})%
  \bibAnnoteFile{NoStop}{Luscombe_PRE_96}%
\bibitem{daSilva_PRE_95}%
  \BibitemOpen
  \bibfield{author}{%
  \bibinfo {author} {\bibfnamefont{J.~K.~L.}\ \bibnamefont{da~Silva}}, \bibinfo
  {author} {\bibfnamefont{A.~G.}\ \bibnamefont{Moreira}}, \bibinfo {author}
  {\bibfnamefont{M.~S.}\ \bibnamefont{Soares}},\ and\ \bibinfo {author}
  {\bibfnamefont{F.~C.~S.}\ \bibnamefont{Barreto}},\ }%
  \bibfield{journal}{%
  \Doi{10.1103/PhysRevE.52.4527}{\bibinfo {journal} {Phys. Rev. E}}\ }%
  \textbf{\bibinfo {volume} {52}},\ \bibinfo {pages} {4527} (\bibinfo {year}
  {1995})%
  \bibAnnoteFile{NoStop}{daSilva_PRE_95}%
\bibitem{PRL_Bogani}%
  \BibitemOpen
  \bibfield{author}{%
  \bibinfo {author} {\bibfnamefont{L.}~\bibnamefont{Bogani}}, \bibinfo {author}
  {\bibfnamefont{A.}~\bibnamefont{Caneschi}}, \bibinfo {author}
  {\bibfnamefont{M.}~\bibnamefont{Fedi}}, \bibinfo {author}
  {\bibfnamefont{D.}~\bibnamefont{Gatteschi}}, \bibinfo {author}
  {\bibfnamefont{M.}~\bibnamefont{Massi}}, \bibinfo {author}
  {\bibfnamefont{M.~A.}\ \bibnamefont{Novak}}, \bibinfo {author}
  {\bibfnamefont{M.~G.}\ \bibnamefont{Pini}}, \bibinfo {author}
  {\bibfnamefont{A.}~\bibnamefont{Rettori}}, \bibinfo {author}
  {\bibfnamefont{R.}~\bibnamefont{Sessoli}},\ and\ \bibinfo {author}
  {\bibfnamefont{A.}~\bibnamefont{Vindigni}},\ }%
  \bibfield{journal}{%
  \Doi{10.1103/PhysRevLett.92.207204}{\bibinfo {journal} {Phys. Rev. Lett.}}\
  }%
  \textbf{\bibinfo {volume} {92}},\ \bibinfo {pages} {207204} (\bibinfo {year}
  {2004})%
  \bibAnnoteFile{NoStop}{PRL_Bogani}%
\bibitem{Coulon04PRB}%
  \BibitemOpen
  \bibfield{author}{%
  \bibinfo {author} {\bibfnamefont{C.}~\bibnamefont{Coulon}}, \bibinfo {author}
  {\bibfnamefont{R.}~\bibnamefont{Cl\'erac}}, \bibinfo {author}
  {\bibfnamefont{L.}~\bibnamefont{Lecren}}, \bibinfo {author}
  {\bibfnamefont{W.}~\bibnamefont{Wernsdorfer}},\ and\ \bibinfo {author}
  {\bibfnamefont{H.}~\bibnamefont{Miyasaka}},\ }%
  \bibfield{journal}{%
  \Doi{10.1103/PhysRevB.69.132408}{\bibinfo {journal} {Phys. Rev. B}}\ }%
  \textbf{\bibinfo {volume} {69}},\ \bibinfo {pages} {132408} (\bibinfo {month}
  {Apr.}\ \bibinfo {year} {2004})%
  \bibAnnoteFile{NoStop}{Coulon04PRB}%
\bibitem{Vindigni_JMMM_04}%
  \BibitemOpen
  \bibfield{author}{%
  \bibinfo {author} {\bibfnamefont{A.}~\bibnamefont{Vindigni}}, \bibinfo
  {author} {\bibfnamefont{L.}~\bibnamefont{Bogani}}, \bibinfo {author}
  {\bibfnamefont{D.}~\bibnamefont{Gatteschi}}, \bibinfo {author}
  {\bibfnamefont{R.}~\bibnamefont{Sessoli}}, \bibinfo {author}
  {\bibfnamefont{A.}~\bibnamefont{Rettori}},\ and\ \bibinfo {author}
  {\bibfnamefont{M.~A.}\ \bibnamefont{Novak}},\ }%
  \bibfield{journal}{%
  \Doi{doi:10.1016/j.jmmm.2003.12.1209}{\bibinfo {journal} {J. Magn. Magn.
  Mater.}}\ }%
  \textbf{\bibinfo {volume} {272-276}},\ \bibinfo {pages} {297} (\bibinfo
  {year} {2004})%
  \bibAnnoteFile{NoStop}{Vindigni_JMMM_04}%
\bibitem{Pini-Rettori_PRB_07}%
  \BibitemOpen
  \bibfield{author}{%
  \bibinfo {author} {\bibfnamefont{M.~G.}\ \bibnamefont{Pini}}\ and\ \bibinfo
  {author} {\bibfnamefont{A.}~\bibnamefont{Rettori}},\ }%
  \bibfield{journal}{%
  \Doi{10.1103/PhysRevB.76.064407}{\bibinfo {journal} {Phys. Rev. B}}\ }%
  \textbf{\bibinfo {volume} {76}},\ \bibinfo {pages} {064407} (\bibinfo {year}
  {2007})%
  \bibAnnoteFile{NoStop}{Pini-Rettori_PRB_07}%
\bibitem{Coulon_PRB_07}%
  \BibitemOpen
  \bibfield{author}{%
  \bibinfo {author} {\bibfnamefont{C.}~\bibnamefont{Coulon}}, \bibinfo {author}
  {\bibfnamefont{R.}~\bibnamefont{Cl\'erac}}, \bibinfo {author}
  {\bibfnamefont{W.}~\bibnamefont{Wernsdorfer}}, \bibinfo {author}
  {\bibfnamefont{T.}~\bibnamefont{Colin}}, \bibinfo {author}
  {\bibfnamefont{A.}~\bibnamefont{Saitoh}}, \bibinfo {author}
  {\bibfnamefont{N.}~\bibnamefont{Motokawa}},\ and\ \bibinfo {author}
  {\bibfnamefont{H.}~\bibnamefont{Miyasaka}},\ }%
  \bibfield{journal}{%
  \Doi{10.1103/PhysRevB.76.214422}{\bibinfo {journal} {Phys. Rev. B}}\ }%
  \textbf{\bibinfo {volume} {76}},\ \bibinfo {pages} {214422} (\bibinfo {year}
  {2007})%
  \bibAnnoteFile{NoStop}{Coulon_PRB_07}%
\bibitem{Pini-Vindigni_JPCM_09}%
  \BibitemOpen
  \bibfield{author}{%
  \bibinfo {author} {\bibfnamefont{A.}~\bibnamefont{Vindigni}}\ and\ \bibinfo
  {author} {\bibfnamefont{M.~G.}\ \bibnamefont{Pini}},\ }%
  \bibfield{journal}{%
  \Doi{10.1088/0953-8984/21/23/236007}{\bibinfo {journal} {J. Phys.: Condens.
  Matter}}\ }%
  \textbf{\bibinfo {volume} {21}},\ \bibinfo {pages} {236007} (\bibinfo {year}
  {2009})%
  \bibAnnoteFile{NoStop}{Pini-Vindigni_JPCM_09}%
\bibitem{Bernot_PRB_09}%
  \BibitemOpen
  \bibfield{author}{%
  \bibinfo {author} {\bibfnamefont{K.}~\bibnamefont{Bernot}}, \bibinfo {author}
  {\bibfnamefont{J.}~\bibnamefont{Luzon}}, \bibinfo {author}
  {\bibfnamefont{A.}~\bibnamefont{Caneschi}}, \bibinfo {author}
  {\bibfnamefont{D.}~\bibnamefont{Gatteschi}}, \bibinfo {author}
  {\bibfnamefont{R.}~\bibnamefont{Sessoli}}, \bibinfo {author}
  {\bibfnamefont{L.}~\bibnamefont{Bogani}}, \bibinfo {author}
  {\bibfnamefont{A.}~\bibnamefont{Vindigni}}, \bibinfo {author}
  {\bibfnamefont{A.}~\bibnamefont{Rettori}},\ and\ \bibinfo {author}
  {\bibfnamefont{M.~G.}\ \bibnamefont{Pini}},\ }%
  \bibfield{journal}{%
  \Doi{10.1103/PhysRevB.79.134419}{\bibinfo {journal} {Phys. Rev. B}}\ }%
  \textbf{\bibinfo {volume} {79}},\ \bibinfo {pages} {134419} (\bibinfo {year}
  {2009})%
  \bibAnnoteFile{NoStop}{Bernot_PRB_09}%
\bibitem{Balanda}%
  \BibitemOpen
  \bibfield{author}{%
  \bibinfo {author} {\bibfnamefont{M.}~\bibnamefont{Balanda}}, \bibinfo
  {author} {\bibfnamefont{M.}~\bibnamefont{Rams}}, \bibinfo {author}
  {\bibfnamefont{S.~K.}\ \bibnamefont{Nayak}}, \bibinfo {author}
  {\bibfnamefont{Z.}~\bibnamefont{Tomkowicz}}, \bibinfo {author}
  {\bibfnamefont{W.}~\bibnamefont{Haase}}, \bibinfo {author}
  {\bibfnamefont{K.}~\bibnamefont{Tomala}},\ and\ \bibinfo {author}
  {\bibfnamefont{J.~V.}\ \bibnamefont{Yakhmi}},\ }%
  \bibfield{journal}{%
  \Doi{10.1103/PhysRevB.74.224421}{\bibinfo {journal} {Phys. Rev. B}}\ }%
  \textbf{\bibinfo {volume} {74}},\ \bibinfo {pages} {224421} (\bibinfo {year}
  {2006})%
  \bibAnnoteFile{NoStop}{Balanda}%
\bibitem{Miyasaka06CEJ}%
  \BibitemOpen
  \bibfield{author}{%
  \bibinfo {author} {\bibfnamefont{H.}~\bibnamefont{Miyasaka}}, \bibinfo
  {author} {\bibfnamefont{T.}~\bibnamefont{Madanbashi}}, \bibinfo {author}
  {\bibfnamefont{K.}~\bibnamefont{Sugimoto}}, \bibinfo {author}
  {\bibfnamefont{Y.}~\bibnamefont{Nakazawa}}, \bibinfo {author}
  {\bibfnamefont{W.}~\bibnamefont{Wernsdorfer}}, \bibinfo {author}
  {\bibfnamefont{K.}~\bibnamefont{ichi Sugiura}}, \bibinfo {author}
  {\bibfnamefont{M.}~\bibnamefont{Yamashita}}, \bibinfo {author}
  {\bibfnamefont{C.}~\bibnamefont{Coulon}},\ and\ \bibinfo {author}
  {\bibfnamefont{R.}~\bibnamefont{Cl\'erac}},\ }%
  \bibfield{journal}{%
  \Doi{10.1002/chem.200600289}{\bibinfo {journal} {Chem. Eur. J.}}\ }%
  \textbf{\bibinfo {volume} {12}},\ \bibinfo {pages} {7028} (\bibinfo {year}
  {2006})%
  \bibAnnoteFile{NoStop}{Miyasaka06CEJ}%
\bibitem{Vindigni08ICA}%
  \BibitemOpen
  \bibfield{author}{%
  \bibinfo {author} {\bibfnamefont{A.}~\bibnamefont{Vindigni}},\ }%
  \bibfield{journal}{%
  \Doi{DOI: 10.1016/j.ica.2008.02.058}{\bibinfo {journal} {Inorg. Chim. Acta}}\
  }%
  \textbf{\bibinfo {volume} {361}},\ \bibinfo {pages} {3731 } (\bibinfo {year}
  {2008})%
  \bibAnnoteFile{NoStop}{Vindigni08ICA}%
\bibitem{Krumhansl75PRB}%
  \BibitemOpen
  \bibfield{author}{%
  \bibinfo {author} {\bibfnamefont{J.~A.}\ \bibnamefont{Krumhansl}}\ and\
  \bibinfo {author} {\bibfnamefont{J.~R.}\ \bibnamefont{Schrieffer}},\ }%
  \bibfield{journal}{%
  \Doi{10.1103/PhysRevB.11.3535}{\bibinfo {journal} {Phys. Rev. B}}\ }%
  \textbf{\bibinfo {volume} {11}},\ \bibinfo {pages} {3535} (\bibinfo {month}
  {May}\ \bibinfo {year} {1975})%
  \bibAnnoteFile{NoStop}{Krumhansl75PRB}%
\bibitem{Nakamura78JPCSSP}%
  \BibitemOpen
  \bibfield{author}{%
  \bibinfo {author} {\bibfnamefont{K.}~\bibnamefont{Nakamura}}\ and\ \bibinfo
  {author} {\bibfnamefont{T.}~\bibnamefont{Sasada}},\ }%
  \bibfield{journal}{%
  \bibinfo {journal} {J. Phys. C: Solid State Phys.}\ }%
  \textbf{\bibinfo {volume} {11}},\ \bibinfo {pages} {331} (\bibinfo {year}
  {1978})%
  \bibAnnoteFile{NoStop}{Nakamura78JPCSSP}%
\bibitem{Fogedby84JPCSSP}%
  \BibitemOpen
  \bibfield{author}{%
  \bibinfo {author} {\bibfnamefont{H.~C.}\ \bibnamefont{Fogedby}}, \bibinfo
  {author} {\bibfnamefont{P.}~\bibnamefont{Hedegard}},\ and\ \bibinfo {author}
  {\bibfnamefont{A.}~\bibnamefont{Svane}},\ }%
  \bibfield{journal}{%
  \bibinfo {journal} {J. Phys. C: Solid State Phys.}\ }%
  \textbf{\bibinfo {volume} {17}},\ \bibinfo {pages} {3475} (\bibinfo {year}
  {1984})%
  \bibAnnoteFile{NoStop}{Fogedby84JPCSSP}%
\bibitem{Seiden}%
  \BibitemOpen
  \bibfield{author}{%
  \bibinfo {author} {\bibfnamefont{J.}~\bibnamefont{Seiden}},\ }%
  \bibfield{journal}{%
  \Doi{10.1051/jphyslet:019830044023094700}{\bibinfo {journal} {J. Phys. Lett.
  (Paris)}}\ }%
  \textbf{\bibinfo {volume} {44}},\ \bibinfo {pages} {947} (\bibinfo {year}
  {1983})%
  \bibAnnoteFile{NoStop}{Seiden}%
\bibitem{Jongh74AP}%
  \BibitemOpen
  \bibfield{author}{%
  \bibinfo {author} {\bibfnamefont{L.~J.}\ \bibnamefont{de~Jongh}}\ and\
  \bibinfo {author} {\bibfnamefont{A.~R.}\ \bibnamefont{Miedema}},\ }%
  \bibfield{journal}{%
  \Doi{10.1080/00018739700101558}{\bibinfo {journal} {Adv. Phys.}}\ }%
  \textbf{\bibinfo {volume} {23}},\ \bibinfo {pages} {1} (\bibinfo {year}
  {1974})%
  \bibAnnoteFile{NoStop}{Jongh74AP}%
\bibitem{Kirschner97}%
  \BibitemOpen
  \bibfield{author}{%
  \bibinfo {author} {\bibfnamefont{J.}~\bibnamefont{Shen}}, \bibinfo {author}
  {\bibfnamefont{R.}~\bibnamefont{Skomski}}, \bibinfo {author}
  {\bibfnamefont{H.~J.}\ \bibnamefont{M.~Klaua}}, \bibinfo {author}
  {\bibfnamefont{S.~S.}\ \bibnamefont{Manoharan}},\ and\ \bibinfo {author}
  {\bibfnamefont{J.}~\bibnamefont{Kirschner}},\ }%
  \bibfield{journal}{%
  \Doi{10.1103/PhysRevB.56.2340}{\bibinfo {journal} {Phys. Rev. B}}\ }%
  \textbf{\bibinfo {volume} {56}},\ \bibinfo {pages} {2340} (\bibinfo {year}
  {1997})%
  \bibAnnoteFile{NoStop}{Kirschner97}%
\bibitem{Tannous}%
  \BibitemOpen
  \bibfield{author}{%
  \bibinfo {author} {\bibfnamefont{R.}~\bibnamefont{Pandit}}\ and\ \bibinfo
  {author} {\bibfnamefont{C.}~\bibnamefont{Tannous}},\ }%
  \bibfield{journal}{%
  \Doi{10.1103/PhysRevB.28.281}{\bibinfo {journal} {Phys. Rev. B}}\ }%
  \textbf{\bibinfo {volume} {28}},\ \bibinfo {pages} {281} (\bibinfo {year}
  {1983})%
  \bibAnnoteFile{NoStop}{Tannous}%
\bibitem{Nature_Gambardella}%
  \BibitemOpen
  \bibfield{author}{%
  \bibinfo {author} {\bibfnamefont{P.}~\bibnamefont{Gambardella}}, \bibinfo
  {author} {\bibfnamefont{A.}~\bibnamefont{Dallmeyer}}, \bibinfo {author}
  {\bibfnamefont{K.}~\bibnamefont{Maiti}}, \bibinfo {author}
  {\bibfnamefont{M.~C.}\ \bibnamefont{Malagoli}}, \bibinfo {author}
  {\bibfnamefont{W.}~\bibnamefont{Eberhardt}}, \bibinfo {author}
  {\bibfnamefont{K.}~\bibnamefont{Kern}},\ and\ \bibinfo {author}
  {\bibfnamefont{C.}~\bibnamefont{Carbone}},\ }%
  \bibfield{journal}{%
  \Doi{10.1038/416301a}{\bibinfo {journal} {Nature}}\ }%
  \textbf{\bibinfo {volume} {416}},\ \bibinfo {pages} {301} (\bibinfo {year}
  {2002})%
  \bibAnnoteFile{NoStop}{Nature_Gambardella}%
\bibitem{Coulon_private}%
  \BibitemOpen
  \bibinfo {note} {Claude Coulon, private communication.}%
  \bibAnnoteFile{Stop}{Coulon_private}%
\bibitem{Polyakov}%
  \BibitemOpen
  \bibfield{author}{%
  \bibinfo {author} {\bibfnamefont{A.~M.}\ \bibnamefont{Polyakov}},\ }%
  \bibfield{journal}{%
  \bibinfo {journal} {Phys. Lett. B}\ }%
  \textbf{\bibinfo {volume} {59}},\ \bibinfo {pages} {79} (\bibinfo {year}
  {1975})%
  \bibAnnoteFile{NoStop}{Polyakov}%
\bibitem{Politi_EPL_94}%
  \BibitemOpen
  \bibfield{author}{%
  \bibinfo {author} {\bibfnamefont{P.}~\bibnamefont{Politi}}, \bibinfo {author}
  {\bibfnamefont{A.}~\bibnamefont{Rettori}}, \bibinfo {author}
  {\bibfnamefont{M.~G.}\ \bibnamefont{Pini}},\ and\ \bibinfo {author}
  {\bibfnamefont{D.}~\bibnamefont{Pescia}},\ }%
  \bibfield{journal}{%
  \bibinfo {journal} {Europhys. Lett.}\ }%
  \textbf{\bibinfo {volume} {28}},\ \bibinfo {pages} {71} (\bibinfo {year}
  {1994})%
  \bibAnnoteFile{NoStop}{Politi_EPL_94}%
\bibitem{Pescia-Pokroksky}%
  \BibitemOpen
  \bibfield{author}{%
  \bibinfo {author} {\bibfnamefont{D.}~\bibnamefont{Pescia}}\ and\ \bibinfo
  {author} {\bibfnamefont{V.~L.}\ \bibnamefont{Pokrovsky}},\ }%
  \bibfield{journal}{%
  \Doi{10.1103/PhysRevLett.65.2599}{\bibinfo {journal} {Phys. Rev. Lett.}}\ }%
  \textbf{\bibinfo {volume} {65}},\ \bibinfo {pages} {2599} (\bibinfo {year}
  {1990})%
  \bibAnnoteFile{NoStop}{Pescia-Pokroksky}%
\bibitem{Garcia-Palacios98PRB}%
  \BibitemOpen
  \bibfield{author}{%
  \bibinfo {author} {\bibfnamefont{J.~L.}\ \bibnamefont{Garcia-Palacios}}\ and\
  \bibinfo {author} {\bibfnamefont{F.~J.}\ \bibnamefont{L\'azaro}},\ }%
  \bibfield{journal}{%
  \bibinfo {journal} {Phys. Rev. B}\ }%
  \textbf{\bibinfo {volume} {58}},\ \bibinfo {pages} {14 937} (\bibinfo {year}
  {1998})%
  \bibAnnoteFile{NoStop}{Garcia-Palacios98PRB}%
\bibitem{Barbara94JMMM}%
  \BibitemOpen
  \bibfield{author}{%
  \bibinfo {author} {\bibfnamefont{B.}~\bibnamefont{Barbara}},\ }%
  \bibfield{journal}{%
  \Doi{DOI: 10.1016/0304-8853(94)90432-4}{\bibinfo {journal} {Journal of
  Magnetism and Magnetic Materials}}\ }%
  \textbf{\bibinfo {volume} {129}},\ \bibinfo {pages} {79 } (\bibinfo {year}
  {1994})%
  \bibAnnoteFile{NoStop}{Barbara94JMMM}%
\bibitem{Taylor20PLMS}%
  \BibitemOpen
  \bibfield{author}{%
  \bibinfo {author} {\bibfnamefont{G.~I.}\ \bibnamefont{Taylor}},\ }%
  \bibfield{journal}{%
  \bibinfo {journal} {Proc. London Math. Soc.}\ }%
  \textbf{\bibinfo {volume} {20}},\ \bibinfo {pages} {196} (\bibinfo {year}
  {1920})%
  \bibAnnoteFile{NoStop}{Taylor20PLMS}%
\bibitem{Brown63PR}%
  \BibitemOpen
  \bibfield{author}{%
  \bibinfo {author} {\bibfnamefont{J.}~\bibnamefont{William Fuller~Brown}},\ }%
  \bibfield{journal}{%
  \bibinfo {journal} {Phys. Rev.}\ }%
  \textbf{\bibinfo {volume} {130}},\ \bibinfo {pages} {1677} (\bibinfo {year}
  {1963})%
  \bibAnnoteFile{NoStop}{Brown63PR}%
\bibitem{Marrows_AdvPhys_05}%
  \BibitemOpen
  \bibfield{author}{%
  \bibinfo {author} {\bibfnamefont{C.~H.}\ \bibnamefont{Marrows}},\ }%
  \bibfield{journal}{%
  \Doi{10.1080/00018730500442209}{\bibinfo {journal} {Adv. Phys.}}\ }%
  \textbf{\bibinfo {volume} {54}},\ \bibinfo {pages} {585} (\bibinfo {year}
  {2005})%
  \bibAnnoteFile{NoStop}{Marrows_AdvPhys_05}%
\bibitem{Thiaville_EPL_05}%
  \BibitemOpen
  \bibfield{author}{%
  \bibinfo {author} {\bibfnamefont{A.}~\bibnamefont{Thiaville}}, \bibinfo
  {author} {\bibfnamefont{Y.}~\bibnamefont{Nakatani}}, \bibinfo {author}
  {\bibfnamefont{J.}~\bibnamefont{Miltat}},\ and\ \bibinfo {author}
  {\bibfnamefont{Y.}~\bibnamefont{Suzuki}},\ }%
  \bibfield{journal}{%
  \bibinfo {journal} {Europhys. Lett.}\ }%
  \textbf{\bibinfo {volume} {69}},\ \bibinfo {pages} {990} (\bibinfo {year}
  {2005})%
  \bibAnnoteFile{NoStop}{Thiaville_EPL_05}%
\bibitem{view_point_Klaeui}%
  \BibitemOpen
  \bibfield{author}{%
  \bibinfo {author} {\bibfnamefont{M.}~\bibnamefont{Kl\"aui}},\ }%
  \bibfield{journal}{%
  \Doi{DOI: 10.1103/Physics.1.17}{\bibinfo {journal} {Physics}}\ }%
  \textbf{\bibinfo {volume} {1}},\ \bibinfo {pages} {17} (\bibinfo {year}
  {2008})%
  \bibAnnoteFile{NoStop}{view_point_Klaeui}%
\bibitem{Nowak_PRB_09}%
  \BibitemOpen
  \bibfield{author}{%
  \bibinfo {author} {\bibfnamefont{C.}~\bibnamefont{Schieback}}, \bibinfo
  {author} {\bibfnamefont{D.}~\bibnamefont{Hinzke}}, \bibinfo {author}
  {\bibfnamefont{M.}~\bibnamefont{Kl\"aui}}, \bibinfo {author}
  {\bibfnamefont{U.}~\bibnamefont{Nowak}},\ and\ \bibinfo {author}
  {\bibfnamefont{P.}~\bibnamefont{Nielaba}},\ }%
  \bibfield{journal}{%
  \Doi{10.1103/PhysRevB.80.214403}{\bibinfo {journal} {Phys. Rev. B}}\ }%
  \textbf{\bibinfo {volume} {80}},\ \bibinfo {pages} {214403} (\bibinfo {year}
  {2009})%
  \bibAnnoteFile{NoStop}{Nowak_PRB_09}%
\bibitem{Ar_Abanov_PRL_10a}%
  \BibitemOpen
  \bibfield{author}{%
  \bibinfo {author} {\bibfnamefont{O.~A.}\ \bibnamefont{Tretiakov}}\ and\
  \bibinfo {author} {\bibfnamefont{A.}~\bibnamefont{Abanov}},\ }%
  \bibfield{journal}{%
  \Doi{10.1103/PhysRevLett.105.157201}{\bibinfo {journal} {Phys. Rev. Lett.}}\
  }%
  \textbf{\bibinfo {volume} {105}},\ \bibinfo {pages} {157201} (\bibinfo {year}
  {2010})%
  \bibAnnoteFile{NoStop}{Ar_Abanov_PRL_10a}%
\bibitem{Ar_Abanov_PRL_10b}%
  \BibitemOpen
  \bibfield{author}{%
  \bibinfo {author} {\bibfnamefont{O.~A.}\ \bibnamefont{Tretiakov}}, \bibinfo
  {author} {\bibfnamefont{Y.}~\bibnamefont{Liu}},\ and\ \bibinfo {author}
  {\bibfnamefont{A.}~\bibnamefont{Abanov}},\ }%
  \bibfield{journal}{%
  \Doi{10.1103/PhysRevLett.105.217203}{\bibinfo {journal} {Phys. Rev. Lett.}}\
  }%
  \textbf{\bibinfo {volume} {105}},\ \bibinfo {pages} {217203} (\bibinfo {year}
  {2010})%
  \bibAnnoteFile{NoStop}{Ar_Abanov_PRL_10b}%
\bibitem{Yamaguchi_APL_05}%
  \BibitemOpen
  \bibfield{author}{%
  \bibinfo {author} {\bibfnamefont{A.}~\bibnamefont{Yamaguchi}}, \bibinfo
  {author} {\bibfnamefont{S.}~\bibnamefont{Nasu}}, \bibinfo {author}
  {\bibfnamefont{H.}~\bibnamefont{Tanigawa}}, \bibinfo {author}
  {\bibfnamefont{T.}~\bibnamefont{Ono}}, \bibinfo {author}
  {\bibfnamefont{K.}~\bibnamefont{Miyake}}, \bibinfo {author}
  {\bibfnamefont{K.}~\bibnamefont{Mibu}},\ and\ \bibinfo {author}
  {\bibfnamefont{T.}~\bibnamefont{Shinjo}},\ }%
  \bibfield{journal}{%
  \Doi{doi:10.1063/1.1847714}{\bibinfo {journal} {Appl. Phys. Lett.}}\ }%
  \textbf{\bibinfo {volume} {86}},\ \bibinfo {pages} {012511} (\bibinfo {year}
  {2005})%
  \bibAnnoteFile{NoStop}{Yamaguchi_APL_05}%
\bibitem{Junginger_APL_07}%
  \BibitemOpen
  \bibfield{author}{%
  \bibinfo {author} {\bibfnamefont{F.}~\bibnamefont{Junginger}}, \bibinfo
  {author} {\bibfnamefont{M.}~\bibnamefont{Kl\"aui}}, \bibinfo {author}
  {\bibfnamefont{D.}~\bibnamefont{Backes}}, \bibinfo {author}
  {\bibfnamefont{U.}~\bibnamefont{R\"udiger}}, \bibinfo {author}
  {\bibfnamefont{T.}~\bibnamefont{Kasama}}, \bibinfo {author}
  {\bibfnamefont{R.~E.}\ \bibnamefont{Dunin-Borkowski}}, \bibinfo {author}
  {\bibfnamefont{L.~J.}\ \bibnamefont{Heyderman}}, \bibinfo {author}
  {\bibfnamefont{C.~A.~F.}\ \bibnamefont{Vaz}},\ and\ \bibinfo {author}
  {\bibfnamefont{J.~A.~C.}\ \bibnamefont{Bland}},\ }%
  \bibfield{journal}{%
  \Doi{doi:10.1063/1.2709989}{\bibinfo {journal} {Appl. Phys. Lett.}}\ }%
  \textbf{\bibinfo {volume} {90}},\ \bibinfo {pages} {132506} (\bibinfo {year}
  {2007})%
  \bibAnnoteFile{NoStop}{Junginger_APL_07}%
\bibitem{Franken_APL_09}%
  \BibitemOpen
  \bibfield{author}{%
  \bibinfo {author} {\bibfnamefont{J.~H.}\ \bibnamefont{Franken}}, \bibinfo
  {author} {\bibfnamefont{P.}~\bibnamefont{M\"ohrke}}, \bibinfo {author}
  {\bibfnamefont{M.}~\bibnamefont{Kl\"aui}}, \bibinfo {author}
  {\bibfnamefont{J.}~\bibnamefont{Rhensius}}, \bibinfo {author}
  {\bibfnamefont{L.~J.}\ \bibnamefont{Heyderman}}, \bibinfo {author}
  {\bibfnamefont{J.-U.}\ \bibnamefont{Thiele}}, \bibinfo {author}
  {\bibfnamefont{H.~J.~M.}\ \bibnamefont{Swagten}}, \bibinfo {author}
  {\bibfnamefont{U.~J.}\ \bibnamefont{Gibson}},\ and\ \bibinfo {author}
  {\bibfnamefont{U.}~\bibnamefont{R\"udiger}},\ }%
  \bibfield{journal}{%
  \Doi{doi:10.1063/1.3265944}{\bibinfo {journal} {Appl. Phys. Lett.}}\ }%
  \textbf{\bibinfo {volume} {95}},\ \bibinfo {pages} {212502} (\bibinfo {year}
  {2009})%
  \bibAnnoteFile{NoStop}{Franken_APL_09}%
\bibitem{Allenspach_PRL_09}%
  \BibitemOpen
  \bibfield{author}{%
  \bibinfo {author} {\bibfnamefont{S.}~\bibnamefont{Lepadatu}}, \bibinfo
  {author} {\bibfnamefont{A.}~\bibnamefont{Vanhaverbeke}}, \bibinfo {author}
  {\bibfnamefont{D.}~\bibnamefont{Atkinson}}, \bibinfo {author}
  {\bibfnamefont{R.}~\bibnamefont{Allenspach}},\ and\ \bibinfo {author}
  {\bibfnamefont{C.~H.}\ \bibnamefont{Marrows}},\ }%
  \bibfield{journal}{%
  \Doi{10.1103/PhysRevLett.102.127203}{\bibinfo {journal} {Phys. Rev. Lett.}}\
  }%
  \textbf{\bibinfo {volume} {102}},\ \bibinfo {pages} {127203} (\bibinfo {year}
  {2009})%
  \bibAnnoteFile{NoStop}{Allenspach_PRL_09}%
\bibitem{Wyld}%
  \BibitemOpen
  \bibfield{author}{%
  \bibinfo {author} {\bibfnamefont{H.~W.}\ \bibnamefont{Wyld}},\ }%
  \emph{\bibinfo {title} {Mathematical Methods of Physics}}\ (\bibinfo
  {publisher} {Benjamin},\ \bibinfo {address} {Massachusetts, USA},\ \bibinfo
  {year} {1976})%
  \bibAnnoteFile{NoStop}{Wyld}%
\bibitem{Stroud}%
  \BibitemOpen
  \bibfield{author}{%
  \bibinfo {author} {\bibfnamefont{A.~H.}\ \bibnamefont{Stroud}},\ }%
  \emph{\bibinfo {title} {Approximate Calculation of Multiple Integrals}}\
  (\bibinfo {publisher} {Prentice-Hall, Englewood Cliffs},\ \bibinfo {address}
  {New Jersey, USA},\ \bibinfo {year} {1971})%
  \bibAnnoteFile{NoStop}{Stroud}%
\bibitem{McLaren}%
  \BibitemOpen
  \bibfield{author}{%
  \bibinfo {author} {\bibfnamefont{A.~D.}\ \bibnamefont{McLaren}},\ }%
  \bibfield{journal}{%
  \bibinfo {journal} {Math. Comp.}\ }%
  \textbf{\bibinfo {volume} {17}},\ \bibinfo {pages} {361} (\bibinfo {year}
  {1963})%
  \bibAnnoteFile{NoStop}{McLaren}%
\bibitem{Abramowitz}%
  \BibitemOpen
  \bibfield{author}{%
  \bibinfo {author} {\bibfnamefont{M.}~\bibnamefont{Abramowitz}}\ and\ \bibinfo
  {author} {\bibfnamefont{I.~E.}\ \bibnamefont{Stegum}},\ }%
  \emph{\bibinfo {title} {Handbook of mathematical functions}}\ (\bibinfo
  {publisher} {Dover},\ \bibinfo {address} {New York, USA},\ \bibinfo {year}
  {1970})%
  \bibAnnoteFile{NoStop}{Abramowitz}%
\bibitem{footnote_Polyakov}%
  \BibitemOpen
  \bibinfo {note} {The number of vectors $\vec{e}_a(x)$ equals the components
  of the slow-varying field $\vec{n}(x)$ minus one. In the present case we have
  two $\vec{e}_a(x)$ vectors while, e.g., in the xy model we would have had
  just one vector $\vec{e}_1(x)$ because $\vec{n}(x)$ is a two-component vector
  field.}%
  \bibAnnoteFile{Stop}{footnote_Polyakov}%
\end{thebibliography}%

\end{document}